\documentclass[usenatbib]{mnras}

\usepackage[T1]{fontenc}
\usepackage{ae,aecompl}

\usepackage{graphicx}	
\usepackage{amsmath}	
\usepackage{amssymb}	
\usepackage{mathrsfs}
\usepackage{enumitem}
\usepackage[normalem]{ulem} 
\usepackage{multirow} 
\usepackage{xcolor} 
\usepackage{lipsum} 
\usepackage{hyperref} 
\usepackage{mathtools} 
\usepackage{subcaption}
\usepackage{array} 
\usepackage{lipsum}
\usepackage{txfonts}
\usepackage{float}
\usepackage{physics}
\usepackage[shortcuts]{extdash}
\usepackage{xcolor}
\DeclareMathOperator*{\argmax}{arg\,max}

\DeclareRobustCommand{\VAN}[3]{#2}
\let\VANthebibliography\thebibliography
\def\thebibliography{\DeclareRobustCommand{\VAN}[3]{##3}\VANthebibliography}

\makeatletter
\def\footnoterule{\kern-3\p@
  \hrule \@width 2in \kern 2.6\p@} 
\makeatother

\title[Cosmic Web Dissection in FDM Cosmologies]{Cosmic Web Dissection in Fuzzy Dark Matter Cosmologies}

\author[T. Dome et al.]{Tibor Dome$^{1,2}$\thanks{E-mail: td448@cam.ac.uk}, Anastasia Fialkov$^{1,2}$, Nina Sartorio$^{3}$, Philip Mocz$^{4}$\\
$^{1}$Institute of Astronomy, University of Cambridge, Madingley Road, Cambridge, CB3 0HA, UK\\
$^{2}$Kavli Institute for Cosmology, Madingley Road, Cambridge, CB3 0HA, UK\\
$^{3}$Sterrenkundig Observatorium, Ghent University, Krĳgslaan 281-S9, B9000 Ghent, Belgium\\
$^{4}$Department of Astrophysical Sciences, Princeton University, 4 Ivy Lane, Princeton, NJ, 08544, USA
}

\date{Accepted XXX. Received YYY; in original form ZZZ}

\pubyear{2023}

\begin{document}
\label{firstpage}
\pagerange{\pageref{firstpage}--\pageref{lastpage}}
\maketitle

\begin{abstract}
On large cosmological scales, anisotropic gravitational collapse is manifest in the dark cosmic web. Its statistical properties are little known for alternative dark matter models such as fuzzy dark matter (FDM). In this work, we assess for the first time the relative importance of cosmic nodes, filaments, walls and voids in a cosmology with primordial small-scale suppression of power. We post-process $N$-body simulations of FDM-like cosmologies with varying axion mass $m$ at redshifts $z\sim 1.0-5.6$ using the NEXUS+ Multiscale Morphology Filter technique at smoothing scale $\Delta x = 0.04 \ h^{-1}$Mpc. The formation of wall and void halos is more suppressed than naively expected from the half-mode mass $M_{1/2}$. Also, we quantify the mass and volume filling fraction of cosmic environments and find that 2D cosmic sheets host a larger share of the matter content of the Universe as $m$ is reduced, with an $\sim 8-12$\% increase for the $m=7 \times 10^{-22}$ eV model compared to CDM. We show that in FDM-like cosmologies, filaments, walls and voids are cleaner and more pronounced structures than in CDM, revealed by a strong mid-range peak in the conditioned overdensity PDFs $P(\delta)$. At high redshift, low-density regions are more suppressed than high-density regions. Furthermore, skewness estimates $S_3$ of the total overdensity PDF in FDM-like cosmologies are consistently higher than in CDM, especially at high redshift $z\sim 5.6$ where the $m=10^{-22}$ eV model differs from CDM by $\sim 6 \sigma$. Accordingly, we advocate for the usage of $P(\delta)$ as a testbed for constraining FDM and other alternative dark matter models.
\end{abstract}

\begin{keywords} cosmology: theory, dark matter, large-scale structure of Universe
\end{keywords}



\section{Introduction}
\label{s_intro}
Far from being uniform, the matter distribution of the Universe on scales\footnote{The Universe is homogeneous and isotropic on scales larger than $\sim 100$ Mpc \citep[e.g.][]{Arjona_2021}. The clearest modern evidence for this \textit{cosmological principle} is in measurements of the Cosmic Microwave Background \citep{Planck_2015}. Its extraordinary uniformity across the sky (with tiny variations of order $\Delta T/T\sim 10^{-5}$) has been coined the \textit{smoothness problem}, typically `resolved' by inflationary cosmology \citep[e.g.][]{Guth_1981}.} of $1-100$ Mpc forms a weblike pattern of voids separated by walls, filaments and nodes. This \textit{cosmic web} \citep{Bond_1996} is still in the linear or quasi-linear phase of collapse and is predicted by the standard model of cosmology \citep{Klypin_1983}. $N$-body simulations (e.g. \cite{Springel_2005, Dolag_2006, Maksimova_2021}) have illustrated how a primordial field of tiny Gaussian density perturbations gives rise to the cosmic web. While the structures that non-linear gravity forms are anisotropic on these scales, the departure from isotropy is only modest. This is suggested by the fact that a small number of multipoles ($l\leq 6$) suffices to capture most of the cosmological information stored in the three-point correlation function (3PCF), as demonstrated by 3PCF estimation techniques \citep{Slepian_2015, Slepian_2016}.\par

The first attempts at mapping the large-scale distribution of galaxies in the universe were made in the late 1970s \citep{Gregory_1978}. \cite{Joeveer_1978} were the first to suggest a resemblance to a cellular system. Since then, a number of large galaxy surveys such as the 2dFGRS \citep{Colless_2003}, the SDSS \citep{Tegmark_2004, Zehavi_2011}, 2MASS \citep{Fairall_2004} and VIPERS \citep{Guzzo_2014} have confirmed the cosmic web.\par

The most prominent features of the cosmic web are filaments, and beyond the well-known Pisces-Perseus chain \citep{Giovanelli_1985} in the local Universe, entire filament inventories have been catalogued \citep[SDSS]{Tempel_2014}. The largest filaments act as highways of the Universe, channelling dark matter (DM), gas and galaxies into the higher density node regions \citep{Knebe_2004, Colberg_2005}. The nodes contain the highest density of galaxy clusters and superclusters such as the Great Attractor \citep{Lynden_Bell_1988}, the Shapley concentration \citep{Ragone_2006} or the Vela supercluster \citep{Kraan_2017}. The 2D components of the cosmic web, sheet-like membranes, are more difficult to detect in the spatial mass distribution traced by galaxies due to their lower surface density. The spatial structure outlined by galaxy clusters, however, does feature flattened supercluster configurations coined \textit{great walls}, the most outstanding of which are the CfA Great Wall \citep{Geller_1989}, the Sloan Great Wall \citep{Gott_2005}, the BOSS Great Wall \citep{Lietzen_2016} and the supergalactic plane \citep{Lahav_2000}. Finally, large void regions are prominent features in redshift surveys as they are practically devoid of any galaxy. Recent studies \citep{Sutter_2012, Leclercq_2015} provide increasingly refined maps and catalogues of the void population in the local Universe. Out of all known voids, the Local Void \citep{Tully_1988} with a diameter of $30$ Mpc is closest to the Milky Way.\par 

Since galaxies are embedded into the cosmic web, their environment codetermines their evolution. In particular, galaxies are often supplied with cold gas via large-scale streams flowing across \textit{intergalactic filaments}, i.e. filaments connecting pairs of neighbouring galaxies \citep{Martin_2016}. Mediated by the embedded halos, cosmic filaments even affect galaxy spin: The spin vectors of small halos tend to align with the axis of filaments they inhabit while those of larger halos tend to be perpendicular to the filament axis. Conceptually, this \textit{spin flip} transition occurring at a halo mass of $\sim 3 \times 10^{12} \ \textup{M}_{\odot}/h$ at redshift zero can be understood using (anisotropic) tidal torque theory \citep{Doroshkevich_1970, Codis_2012, Codis_2015} which predicts a quadrupolar pattern in the vorticity field around filaments.\par 

Recently, it has been found in both simulations \citep{Xia_2021} and observations \citep[SDSS]{Wang_2021} that net rotations of cosmic filaments themselves are non-zero, making them the largest known structures in the Universe to rotate. Cosmic filaments also host the majority of the baryons at low redshifts in the form of the warm-hot intergalactic medium, thereby potentially solving the so-called ``missing baryon problem'' \citep{Klar_2012, Yang_2022, ChavesMontero_2021}.\par

Though only observable via tracer fields, the cosmic web, and thereby the configuration and abundance of each of its morphological components, is highly sensitive to microscopic DM properties. Many aspects of the large-scale structure of the Universe such as the Cosmic Microwave Background \citep[CMB]{Planck_2015}, the Lyman-$\alpha$ forest \citep{Irsic_2016}, galaxy clustering \citep{Nuza_2013} and weak gravitational lensing \citep{Murata_2018} can be described successfully using cosmological models in which $27 \pm 1$\% of the critical mass-energy density consists of cold dark matter (CDM). However, CDM encounters many difficulties when modelling structures on scales of $\sim10$ kpc or less. Two problems among many are the discrepancy between galaxy density profiles in CDM models and observations (the ``core-cusp'' problem \citep{Blok_2010}) and the fact that while classical bulges should be commonplace after major mergers predicted by CDM, about 80\% of observed galaxies exhibit no such bulges \citep{Governato_2010}. Typically, the lack of success is explained by the difficulty of modelling baryonic physics such as star formation \citep{Okamoto_2008}, supernovae \citep{Mashchenko_2008} and black-hole feedback \citep{Delliou_2011}. See \cite{Popolo_2017} for a review on attempts at (unified) baryonic solutions to the $\Lambda$CDM small-scale problems.\par

Alternatively, these problems could be solved by considering distinct types of DM. One DM model that has gained increasing attention is fuzzy dark matter \citep[FDM,][]{Hui_2017, Hu_2000}. This model assumes DM is comprised of extremely light bosons ($m \sim 10^{-22}$ eV) having a de Broglie wavelength $\lambda_{\text{dB}} \sim 1$ kpc. Motivations for the existence of multiple species of light axions are ample and range from well-established predictions of string/M-theory \citep{Arvanitaki_2010, Demirtas_2018} to various field theory extensions \citep{Peccei_1977, Kim_2016} of the standard model. The largely redshift-insensitive comoving de Broglie wavelength $\lambda_{\text{db,c}} \sim (1 + z)^{1/4}m^{-1/2}$ simultaneously suppresses small-scale structure and limits the central density of collapsed halos \citep{Schive_2014}, naturally solving some of the small-scale challenges present in CDM. A distinctive feature of FDM are solitonic cores in the centres of halos, which in models with axion self-interactions can undergo a phase transition from dilute to dense ones \citep{Mocz_2023}. In this work, we look at DM models with a cutoff in the primordial power spectrum, focusing on axion masses in the range $m\in [10^{-22}, \ 2\times 10^{-21}]$ eV. On the scales of interest, this model serves as a proxy for FDM (see Sec. \ref{ss_sims}) and other DM scenarios with a small-scale cutoff such as warm dark matter \citep[WDM,][]{Paduroiu_2022}.\par

In this work, we focus on intermediate scales with wavenumbers $k \sim 0.16 - 80 \ h \, \text{Mpc}^{-1}$ (corresponding to a box of side length $L_{\text{box}} = 40\ h^{-1}$Mpc and resolution $N=1024^3$) at redshifts $z\sim 1.0-5.6$ and quantify the impact of a small-scale cutoff in the primordial power spectrum on the statistics of the cosmic web using cosmological $N$-body simulations. In the more pristine and quasi-linear high-redshift cosmic web, a primordial small-scale cutoff is manifested stronger than in the increasingly non-linear low-redshift Universe where small-scale power is replenished with power from larger scales \citep{Viel_2012}.\par 

The organisation of the paper is as follows: In Sec. \ref{ss_sims}, we describe our large-scale simulations. In Sec. \ref{ss_nexus}, we briefly summarise our independent implementation of the NEXUS+ algorithm \citep{Cautun_2012}, a multiscale morphological analysis tool that identifies all the cosmic structures in a scale-free way. We introduce the concept of cosmic skewness in Sec. \ref{ss_skewness}. Results for the high-$z$ statistics of the cosmic web in FDM-like cosmologies are given in Sec. \ref{s_stats}. We discuss our main findings in Sec. \ref{s_conclusions}. In Appendix \ref{s_vir}, we provide some background on quasi-virialised cosmic filaments while Appendix \ref{s_res_tests} is dedicated to `convergence' tests as a function of resolution scale.

\section{Numerical Methods}
\label{s_num_methods}
\subsection{CDM and FDM-Like Simulations}
\label{ss_sims}
We search for statistical differences in the large-scale morphological components using cosmological $N$-body simulations performed with the state-of-the-art code \scshape{arepo}  \normalfont described by \cite{Arepo_2010} and \cite{Weinberger_2020}. Gravitational forces are computed using a TreePM method \citep{Bagla_2002} which accelerates long-range force calculations by performing them on a particle mesh and short-range ones by hierarchical organisation using a tree-like multipole expansion scheme.\par

\begin{figure*}
\hspace{0.2cm}
\begin{subfigure}{0.49\textwidth}
\includegraphics[scale=0.8]{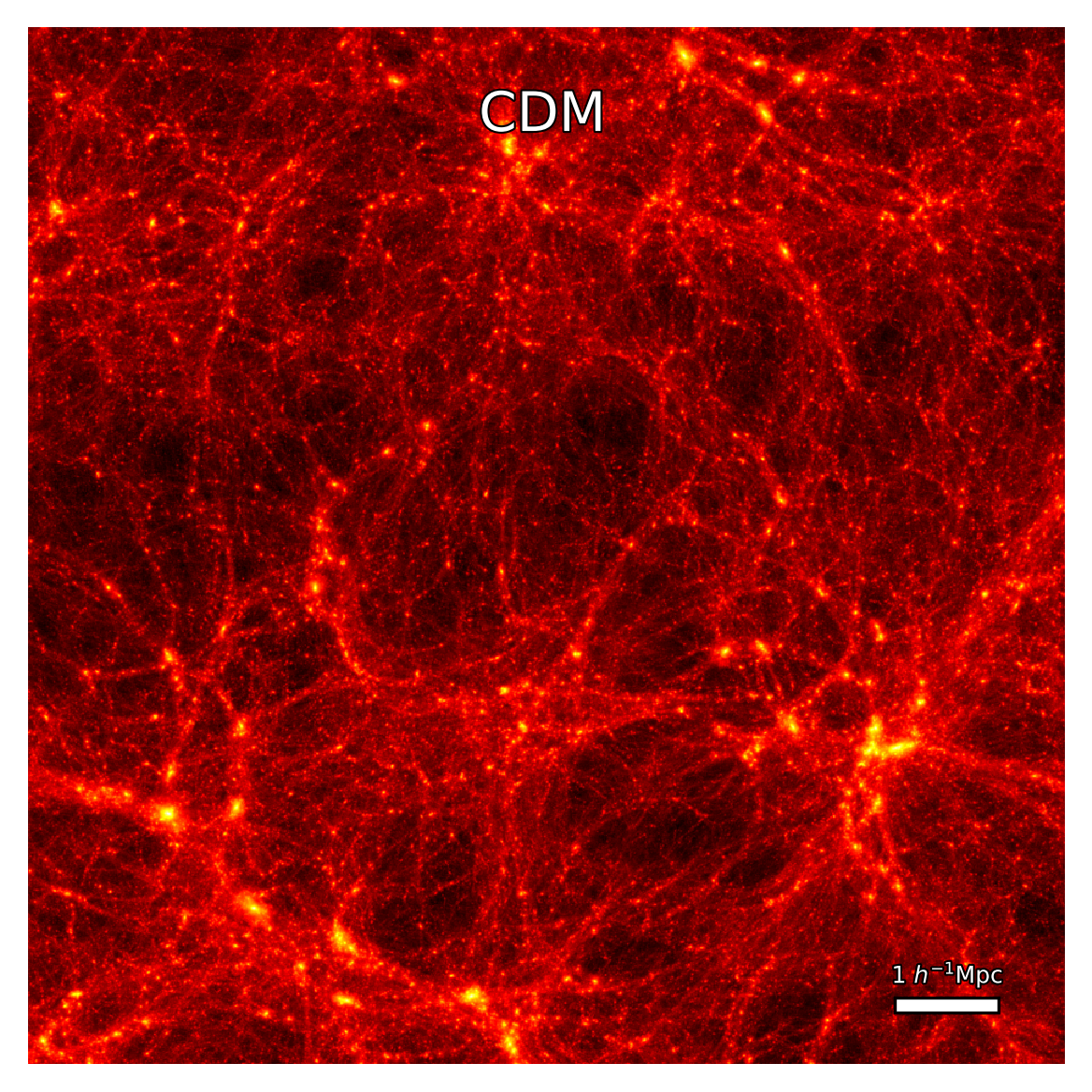}
\end{subfigure}
\begin{subfigure}{0.49\textwidth}
\includegraphics[scale=0.8]{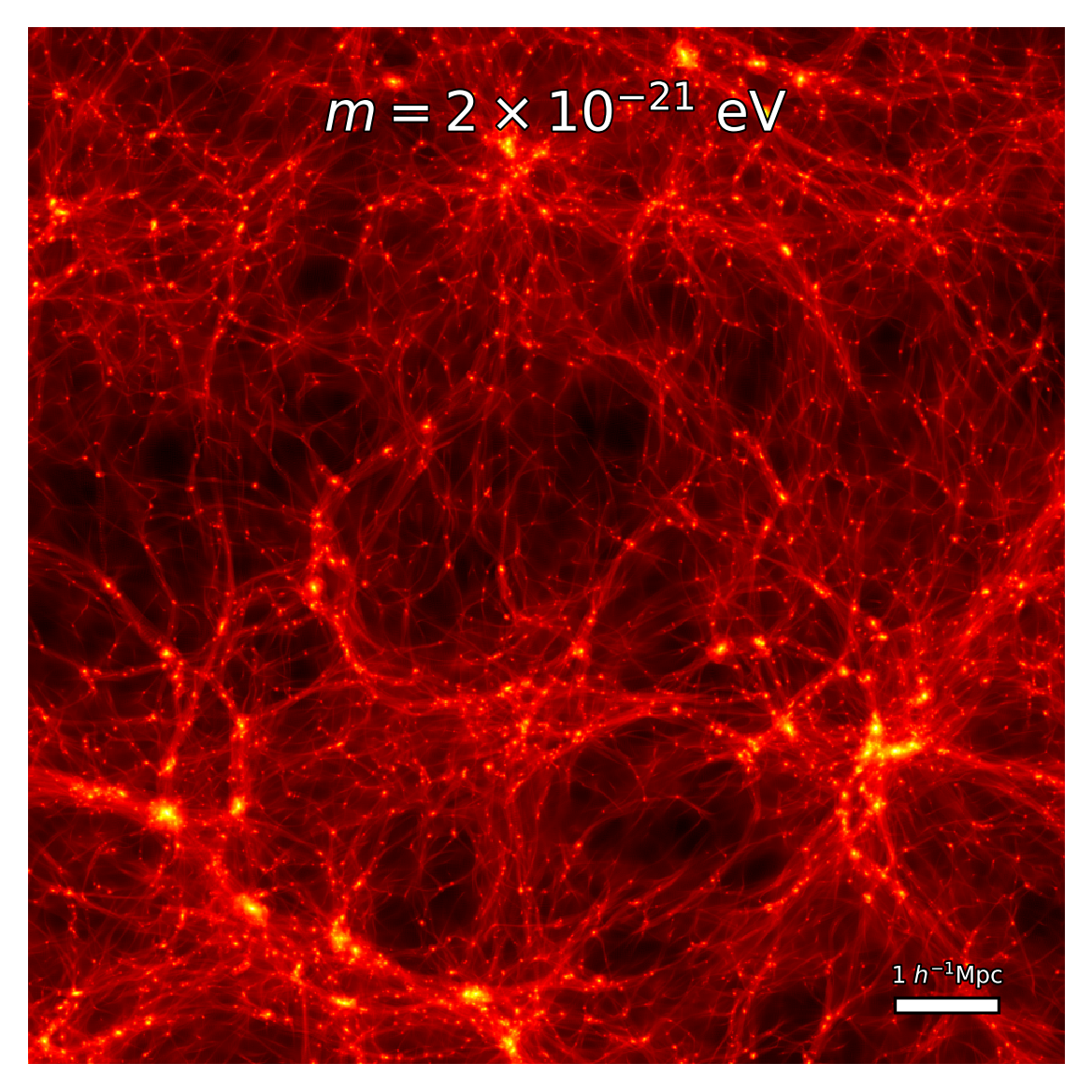}
\end{subfigure}
\begin{subfigure}{0.49\textwidth}
\hspace{0.1cm}
\includegraphics[scale=0.8]{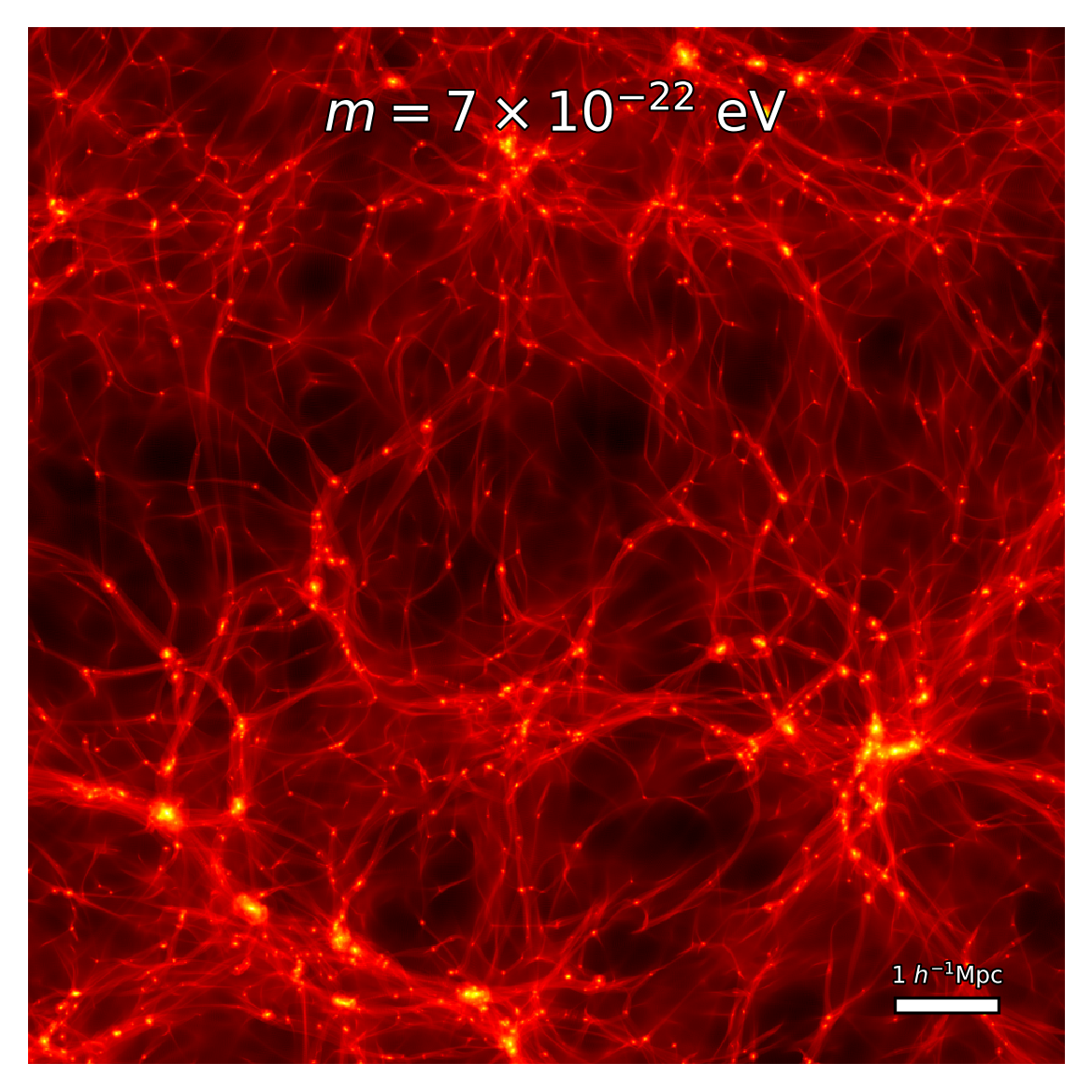}
\end{subfigure}
\begin{subfigure}{0.49\textwidth}
\hspace{0.1cm}
\includegraphics[scale=0.8]{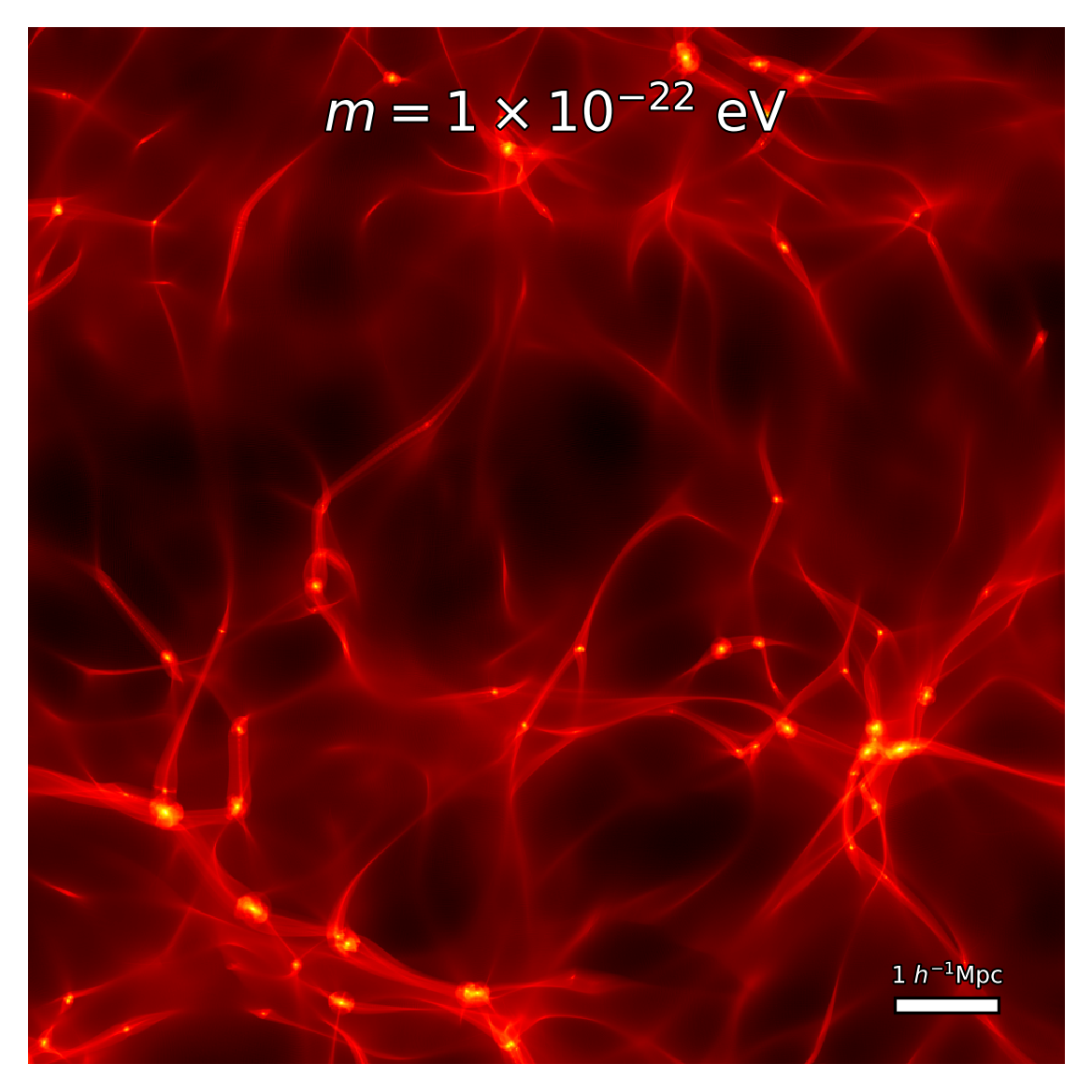}
\end{subfigure}
\caption{Comoving DM density field in a $9 \times 9 \times 9 \ (h^{-1}\text{Mpc})^3$ volume of the $N$-body CDM (top left) and cFDM ($m=2\times 10^{-21}$ eV run: top right, $m=7\times 10^{-22}$ eV run: bottom left and $m=10^{-22}$ eV run: bottom right) runs with $1024^3$ resolution and $L_{\text{box}} = 10\ h^{-1}$Mpc at $z=3.9$, shown in logarithmic projection.}
\label{f_cic_plot}
\end{figure*}

FDM simulations are significantly more challenging than CDM simulations as in the densest regions the wavelike matter oscillations can attain high frequencies $\omega \propto m^{-1}\lambda_{\text{dB}}^{-2}$, requiring very fine temporal resolution even for moderate spatial resolution. To bypass the challenges of FDM simulations, here we employ FDM-like modeling described in \cite{Dome_2022}. In short, we impose a cutoff in the primordial power spectrum similar to \cite{Ni_2019}. This cutoff serves as a proxy for axions generated via vacuum realignment assuming gravitational interactions do not re-thermalise axions.\par

In \cite{Dome_2022}, this proxy for FDM has been termed classical FDM (cFDM). For scales corresponding to wavenumbers $k \sim 0.16 - 80 \ h \, \text{Mpc}^{-1}$, which we explore here, the dynamical manifestation of FDM - the \textit{quantum pressure} \citep[fluid formulation,][]{Madelung_1927} - has only small impact on the growth of DM fluctuations. Specifically, the absolute fractional difference between growth rates in FDM vs cFDM is less than $5$\% for particle masses around $m\sim 10^{-22}$ eV and (halo) mass scales around $M \sim 4\times 10^9 \ M_{\odot}/h$ \citep[see e.g.][]{Corasaniti_2017}. As opposed to a superfluid, cFDM approximates FDM as a classical collisionless fluid, governed by the Vlasov-Poisson system of equations, but with FDM initial conditions. The exponential-like small-scale suppression in the primordial power spectrum which we will often refer to as a cutoff is modelled using the Boltzmann solver \scshape{AxionCamb}\normalfont \ \citep{Hlozek_2015}.

$N$-body simulations offer a much cleaner platform to isolate the imprints of a cutoff in the initial power spectrum on the cosmic web, as in full hydrodynamical runs those imprints are entangled with resolution effects that are due to baryonic physics \citep[e.g.][]{Vogelsberger_2013, Chua_2019}. In our $N$-body suite, we use cosmological volumes with two different box side lengths, $L_{\text{box}} = 10$ and $40 \ h^{-1}$Mpc, for each of three DM resolutions, $N=256^3$, $512^3$, and $1024^3$. To provide a visual understanding of the performance of the NEXUS+ algorithm in detecting key features, we use the set of simulations with $L_{\text{box}} = 10 \ h^{-1}$Mpc while for the bulk of this work we have $L_{\text{box}} = 40 \ h^{-1}$Mpc. The latter box size better balances the competing demands of high resolution of the cosmic web and large volume (to obtain accurate statistical distributions).\par 

Apart from CDM, we run cFDM simulations over a range of axion masses $m=10^{-22}, \ 7\times 10^{-22}, \ 2\times 10^{-21}$ eV. DM halos are identified using the friends-of-friends ({\fontfamily{cmtt}\selectfont FoF}) algorithm with a standard linking length of $b = 0.2 \times \text{(mean inter-particle separation)}$ \citep{Springel_2001_2}. As a minimum halo resolution, we require all halos to be composed of at least $200$ DM resolution elements, i.e. $M_{\text{min}} = 200 \times m_{\mathrm{DM}} = 10^9 \ h^{-1} M_{\odot}$ for the runs with $1024^3$ resolution, $L_{\text{box}} = 40 \ h^{-1}$Mpc. We adopt a Planck cosmology \citep{Planck_2015} with $\Omega_m = 0.3089$, $\Omega_{\Lambda} = 0.6911$, $h=H_0/100=0.6774$ and $\sigma_8 = 0.8159$. Initial conditions are set up at $z=127$, using $n_s = 0.9665$ for the primordial power spectrum of CDM and as input to \scshape{AxionCamb}\normalfont.\par 

Density field projections in various cosmologies are shown in Fig. \ref{f_cic_plot}, for the exemplary redshift of $z=3.9$. On scales much larger than the local de-Broglie wavelength $\lambda_{\text{dB}}$, CDM can be thought of as a limiting case of FDM in the following sense: The FDM potential field converges to the classical answer as $\mathcal{O}(\hbar/m)^2$, thus any superfluid dynamics recovers the classical collisionless limit as $\hbar/m \rightarrow 0$, even in the case of multi-stream flows. However, the density field fails to converge due to order unity fluctuations driven by interference and the uncertainty principle \citep{Mocz_2018}. Mathematically, there is no exact correspondence between the Schr\"odinger-Poisson and the Vlasov-Poisson equation since the former describes a fluid while the latter collisionless particles. FDM remains a fluid even for $\hbar/m \rightarrow 0$. This quasi-correspondence between FDM and CDM in the large-$m$ limit, or rather, the \textit{exact} correspondence between cFDM and CDM, is visible in the density field projections of Fig. \ref{f_cic_plot}. They illustrate how small-scale power is suppressed as the axion mass is reduced from infinity (CDM, top left) all the way down to $m=10^{-22} \ \text{eV}$ (bottom right).

\subsection{Cosmic Web Segmentation: NEXUS+}
\label{ss_nexus}

While (quasi-)virialised filaments as described in Appendix \ref{s_vir} are a useful theoretical concept, to the best of our knowledge none of the cosmic filament finders are based on any virialisation condition. The main reason is that many filaments are far from being virialised. Such non-virialised filaments deviate greatly from the analytical estimates in Eqs. \eqref{Plummer}, \eqref{VirTheoremFils} and \eqref{e_steady_state_sheet_approx} and often constitute high-redshift predecessors of virialised filaments. Since the imposition of a virialisation condition would lead to many prominent structures being missed, we refrain from doing so in the following, as is common.\par

Beyond the ambiguities in the definition of its components, mapping the cosmic web in simulations is a non-trivial task since it lacks structural symmetries that would simplify its analytical treatment. Also, even though the cosmic web is \textit{not} a fractal\footnote{Notably, a fractal model with no transition to homogeneity, as Mandelbrot proposed \cite{Mandelbrot_1983}, is in conflict with the standard FLRW cosmology. See \cite{Gaite_2019} on how to apply fractal models to the cosmic web.} with a clear-cut fractal index, its four components, nodes, filaments, walls and voids, can occur over vast spatial scales. Finally, there is a very wide range of densities found in the cosmic matter distribution, which can spuriously extend/compress the identified component beyond/below its ``natural'' boundary.\par 

\begin{figure*}
\hspace{0.2cm}
\begin{subfigure}{0.49\textwidth}
\includegraphics[scale=0.8]{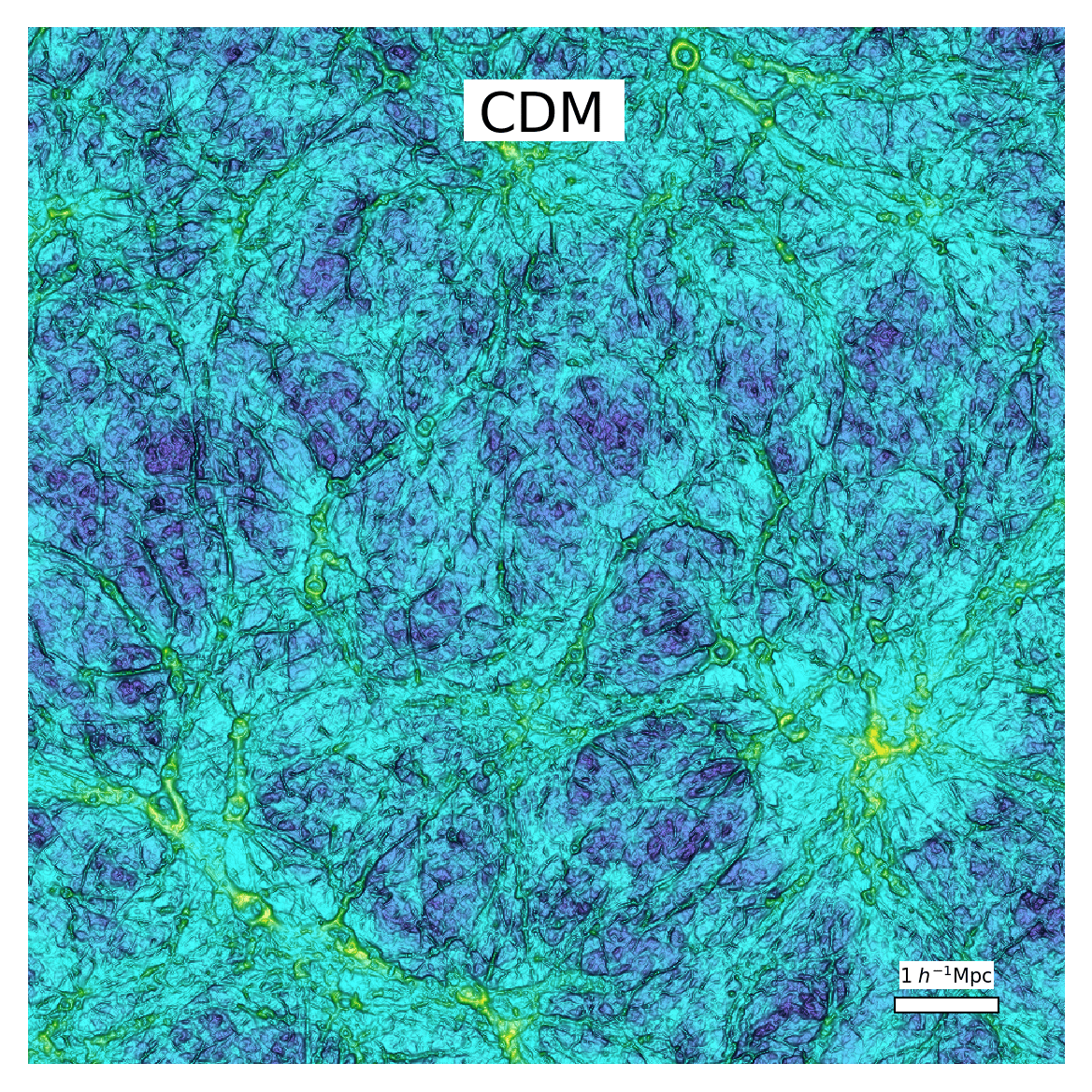}
\end{subfigure}
\begin{subfigure}{0.49\textwidth}
\includegraphics[scale=0.8]{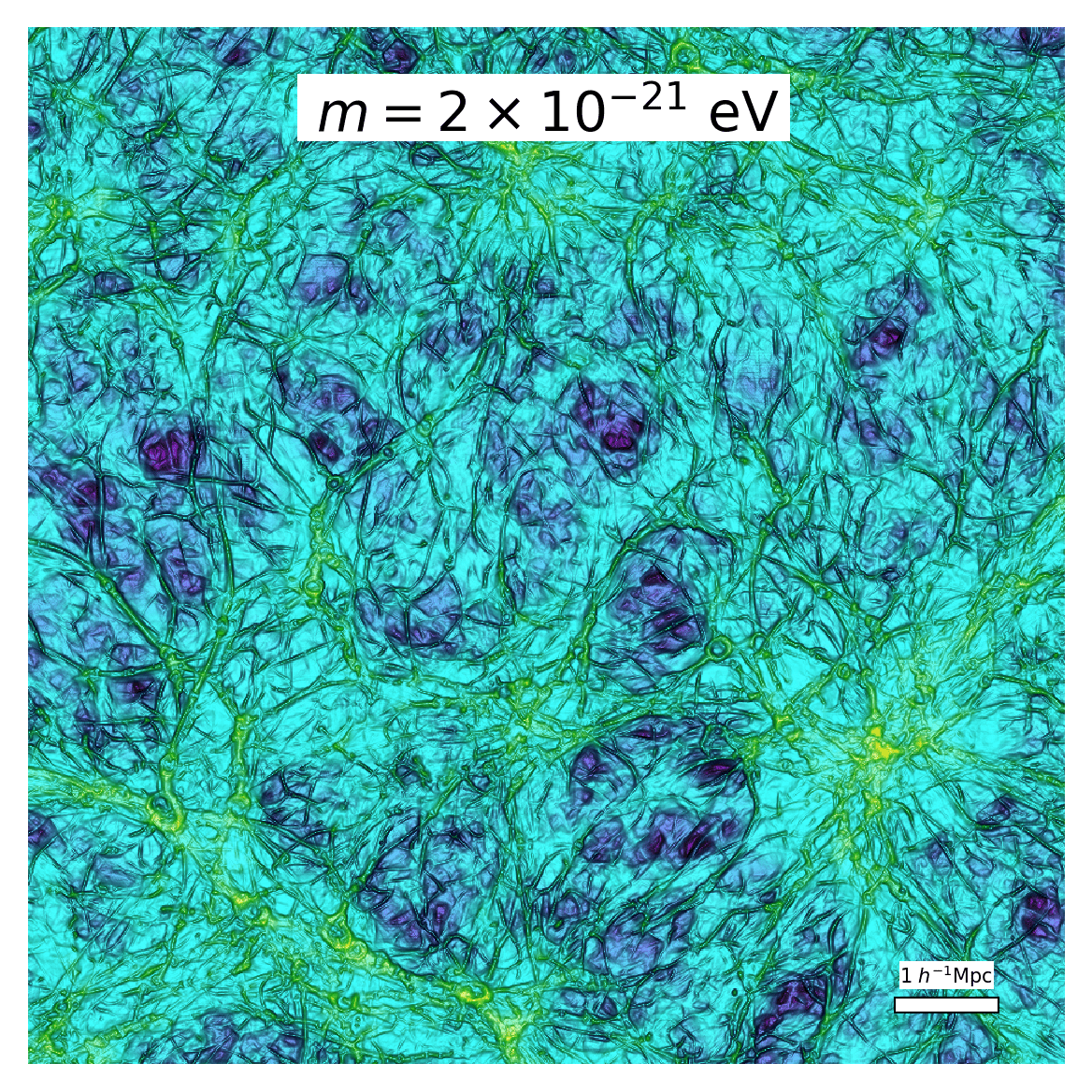}
\end{subfigure}
\begin{subfigure}{0.49\textwidth}
\hspace{0.1cm}
\includegraphics[scale=0.8]{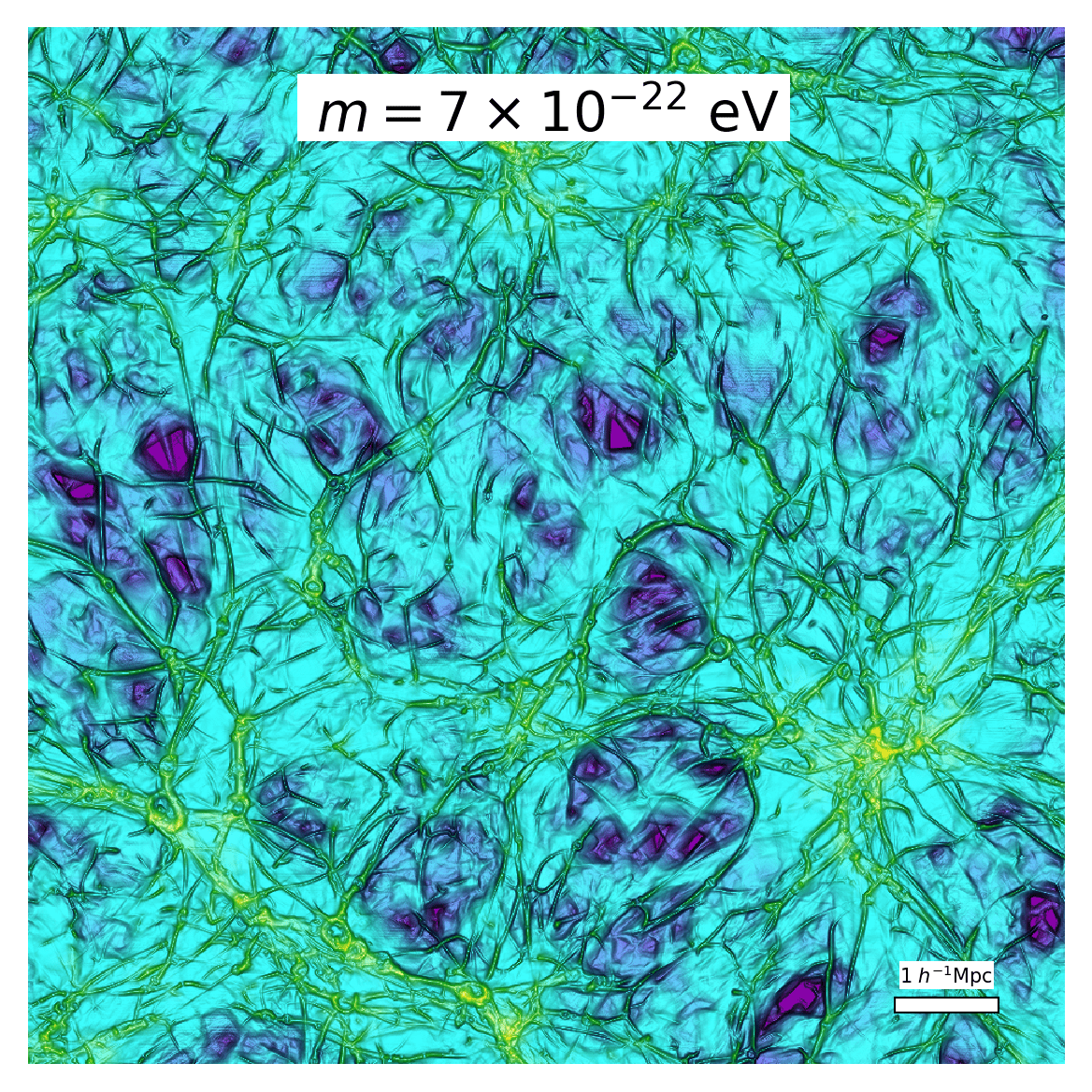}
\end{subfigure}
\begin{subfigure}{0.49\textwidth}
\hspace{0.1cm}
\includegraphics[scale=0.8]{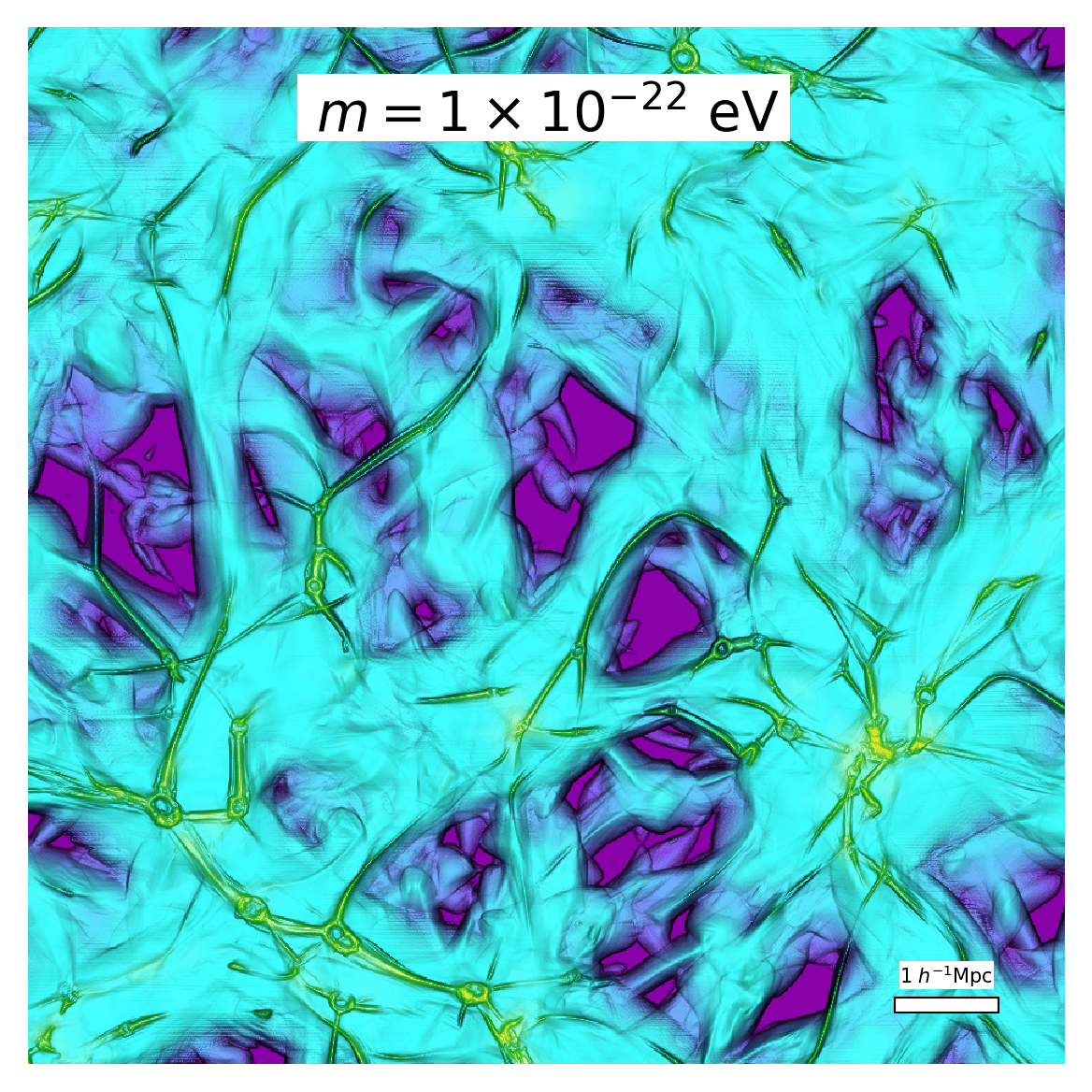}
\end{subfigure}
\caption{An illuminated 2D projection of the filamentary NEXUS+ network $\mathcal{S}_f(\mathbf{x})$ after scale-space stacking and imposition of $\mathcal{S}_{f,\mathrm{cut}}$ in logarithmic projection at redshift $z=3.9$ in a $9 \times 9 \times 9 \ (h^{-1}\text{Mpc})^3$ volume (corresponding to Fig. \ref{f_cic_plot}) of the $N$-body CDM (top left) and cFDM runs ($m=2\times 10^{-21}$ eV run: top right, $m=7\times 10^{-22}$ eV run: bottom left and $m=10^{-22}$ eV run: bottom right) with $1024^3$ resolution and $L_{\text{box}} = 10 \ h^{-1}$Mpc.}
\label{f_fils_viz}
\end{figure*}

Cosmic web detection algorithms can be categorised into graph- and percolation-based techniques \citep{Shandarin_2010, Alpaslan_2014, Naidoo_2022}, stochastic methods \citep{Stoica_2010, Genovese_2010}, topological methods \citep{Aragon_2010} and Hessian-based approaches, see \cite{Libeskind_2017} for a comparison. Our focus in the following is on Hessian-based approaches which exploit morphological information in the gradient (or Hessian) of the density, tidal or velocity shear fields. Most of the Hessian-based formalisms are defined on one particular smoothing scale for the field involved \citep{Hahn_2007, Bond_2010}, but in the last decade explicit multiscale versions have been developed, including \citet{Aragon_2017}.\par 

In this work we use our own implementation of NEXUS+, which we briefly summarize here. NEXUS+ is likewise a morphological multiscale approach based on the density field Hessian. By log-smoothing the density field over a range of spatial filter sizes $R_n \in (R_0, ..., R_N)$ and maximising over the signatures in this so-called \textit{scale-space}, it detects at which scales and locations the various morphological signatures are most prominent. The six steps of the algorithm along with our implementation choices are as follows.
\begin{enumerate}
\item \label{step_i} \textit{Applying a log-Gaussian filter of width $R_n$ to the input field}. If the input field is continuous and denoted by $ f(\vec{x}) $, smoothing the log-field $g = \log_{10} f$ amounts to a simple multiplication with the Gaussian exponential in Fourier space:

\begin{equation}
g_{R_n}(\vec{x}) = \int_{\mathbb{R}^3} \frac{d^3\vec{k}}{(2\pi)^3}e^{-\vec{k}^2\frac{R_n^2}{2}}\hat{g}(\vec{k})e^{i\vec{k}\vec{x}}.
\end{equation}

This yields the NEXUS+ smoothed field $f_{R_n}(\vec{x}) = C_{R_n}10^{\mbox{\footnotesize$g_{R_n}(\vec{x})$}}$ by exponentiation, where $C_{R_n}$ assures the mean of the input field is the same before and after filtering.

\item \textit{Computing Hessian eigenvalues.} The Fourier transform of the Hessian $H_{ij,R_n}(\vec{x})=R_n^2\frac{\partial^2 f_{R_n}(\vec{x})}{\partial x_i \partial x_j}$ reads
\begin{equation}
H_{ij,R_n}(\vec{k})=-k_ik_jR_n^2\hat{f}_{R_n}(\vec{k}).
\end{equation} 

\item \label{step_iii}\textit{Assigning to each point a cluster, filament and wall signature.} The three eigenvalues $\lambda_1 \leq \lambda_2 \leq \lambda_3$ of the Hessian $H_{ij,R_n}(\vec{x})$ can be combined into a shape strength as
\begin{equation}
\mathcal{I}_{R_n} = 
\begin{cases}
	\ \big|\frac{\lambda_3}{\lambda_1} \big| & \text{node} \\ 
	\ \big|\frac{\lambda_2}{\lambda_1} \big| \Theta\left(1-\big|\frac{\lambda_3}{\lambda_1} \big|\right) & \text{filament}\\
	\ \Theta\left(1-\big|\frac{\lambda_2}{\lambda_1} \big|\right) \Theta\left(1-\big|\frac{\lambda_3}{\lambda_1} \big|\right) & \text{wall},
\end{cases}
\end{equation}
where we use the notation $\Theta(x) = x\theta(x)$ for clarity, with $\theta(x)$ the Heaviside step function ($\theta(x) = 1  \ \text{if} \ x \geq 0, 0$ otherwise). We thus obtain the cluster/filament/wall signature as:
\begin{equation}
\mathcal{S}_{R_n} = \mathcal{I}_{R_n}\times
\begin{cases}
	\ |\lambda_3|\theta(-\lambda_1)\theta(-\lambda_2)\theta(-\lambda_3) & \text{node} \\ 
	\ |\lambda_2|\theta(-\lambda_1)\theta(-\lambda_2) & \text{filament}\\
	\ |\lambda_1|\theta(-\lambda_1) & \text{wall}.
\end{cases}
\end{equation}
We multiply by $|\lambda_i|$ to distinguish between noise (small $|\lambda_i|$) and real signals (large $|\lambda_i|$) while the $\theta(-\lambda_i)$ factors impose the necessary eigenvalue constraints.

\item \label{step_iv}
\textit{Computing the environmental signature over a range of smoothing scales}. We repeat steps \ref{step_i}-\ref{step_iii} over a range of smoothing scales ($R_0$, $R_1$, ..., $R_N$). While NEXUS+ is a multi-scale approach, the hierarchy of smoothing scales is a user input. In view of computational feasibility, we opt for relative $\sqrt{2}$-spacings following \cite{Cautun_2012}, i.e. $R_n = (\sqrt{2})^nR_0$, where $R_0$ is the smallest scale at which to expect to find structures. We comment on $R_0$ and the discretization of $ f(\vec{x}) $ below while $N$ is chosen such that $R_N$ does not exceed $4 \ h^{-1}$Mpc. At higher redshift of $z>1$, smaller values for $R_N$ would suffice.

\item \textit{Scale-space stacking.} The scale-independent map is constructed by taking the maximum signature over all scales
\begin{equation}
\mathcal{S}(\vec{x}) = \max_{\text{levels } n} \mathcal{S}_{R_n}(\vec{x}),
\end{equation}
which characterizes the degree to which each voxel $\vec{x}$ is part of a cluster, filament or wall.

\item \textit{Computing the detection threshold}. As the last step, we impose physical criteria to determine the detection threshold corresponding to valid environments. For nodes, the threshold signature $\mathcal{S}_{c,\mathrm{cut}}$ is found by requiring that at least half of the connected regions are virialised according to Eq. \eqref{e_bryan}. This is in contrast to the original papers \citep{Cautun_2012, Cautun_2014}, where the authors use a virialisation overdensity of $\Delta_{\text{vir}} = 370$. To identify connected regions for each node signature floor $\mathcal{S}_{c}$, we label them based on a $1$-connectivity neighbourhood condition. The value $1$ refers to the maximum number of orthogonal hops to consider a voxel a neighbour.\par

Voxels that do not pass this threshold are assigned a node signature of zero. Voxels that do pass the threshold constitute genuine node structures, and after setting the real-space density values at their location to the mean density of the Universe \cite[rather than zero,][]{Cautun_2012}, the slightly modified input field $\tilde{\delta}(\mathbf{x})$ becomes the basis for the calculation of filament signatures $\mathcal{S}_{f,R_n}(\mathbf{x})$ in \ref{step_iii}. The procedure is similar to the one for nodes, except that for filaments, the threshold signature is determined by calculating the mass $M_f(\mathcal{S}_f)$ in filaments with a signature value larger or equal to $\mathcal{S}_f$ and maximizing the mass change with signature:
\begin{equation}
\label{e_fw_det}
\argmax_{\mathcal{S}_f} \ \bigg\lvert \frac{\mathrm{d}M_f^2}{\mathrm{d}\log \mathcal{S}_f}\bigg\rvert = \mathcal{S}_{f,\mathrm{cut}}.
\end{equation}
After identifying filaments and setting the real-space density values at their location to the mean density of the Universe, we identify walls using the same detection threshold \eqref{e_fw_det}. The remaining voxels are automatically identified as voids, which thus constitute the complement to nodes, filaments and walls.
\end{enumerate}

Due to its scale-space approach, NEXUS+ is better equipped to reveal tenuous cosmic environments than classification schemes based on a single scale dissection \citep{Hahn_2007, Bond_2010}. To obtain a complete census of cosmic web environments for the lowest-mass halos and by extension the faintest galaxies, tenuous tendril-like filaments criss-crossing the underdense regions become important. NEXUS+ is admittedly less reliable in the detection of cosmic web nodes than the NEXUS\textunderscore den or NEXUS\textunderscore tidal methods, which are other realisations of the NEXUS algorithm. Shape and location of NEXUS+ node boundaries are oversensitive to the substructure at the periphery \citep{Cautun_2012}. However, for the sake of simplicity and comparing our results to previous works \citep{Cautun_2014, Hellwing_2021}, we use NEXUS+ to identify all cosmic web environments: nodes, filaments, walls and voids.\par 

Rather than the mass-weighted Lagrangian Delaunay tessellation field estimator (DTFE), we choose the volume-weighted Eulerian cloud-in-cell (CIC) mass density field $\delta(\mathbf{x})$ as input to the NEXUS+ algorithm. The regular grid spacing of the input field equals the resolution scale $\Delta x = L_{\text{box}}/N_{\text{lin}} = 0.04 \ h^{-1}$Mpc. The smallest smoothing scale $R_0$ in step \ref{step_iv} is chosen to be twice the grid spacing.\par

For the above set of parameter choices and definition of signature thresholds, Fig. \ref{f_fils_viz} shows a 2D projection of the filament signature field $\mathcal{S}_f(\mathbf{x})$ after scale-space stacking and after imposition of the threshold signature $\mathcal{S}_{f,\mathrm{cut}}$. The intricate filigree of filaments surrounded by vast empty regions is well discernible in Fig. \ref{f_fils_viz}, as well as filamentary signatures being zero at the location of cosmic nodes. This is expected since each voxel receives only one labelling: node, filament, wall or void. As the axion mass $m$ is gradually reduced, the primordial power spectrum cutoff migrates to a larger spatial scale. Concomitant with this migration, we observe a gradual removal of the thinnest filaments. In addition, we find that filaments become visually smoother, which is related to smoother DM accretion onto halos \citep{Khimey_2021}.\par

\subsection{Skewness Analysis}
\label{ss_skewness}
Given a cosmological model, one of the central theoretical goals is to predict the distribution of the cosmic matter overdensity
\begin{equation}
\delta = \frac{\rho-\bar{\rho}}{\bar{\rho}},
\end{equation}
where $\rho$ denotes the matter density and $\bar{\rho} = \rho_{\mathrm{crit}}\Omega_m = 3H_0^2\Omega_m/(8\pi G)$ is the mean matter density of the Universe. The properties of the distribution can be studied analytically using cosmological perturbation theory \citep[PT,][]{Peebles_1980, Bernardeau_1995, Bernardeau_2002} or, as in our case, numerically using $N$-body simulations \citep{Kofman_1994, Mao_2014, Shin_2017, Hellwing_2020}. The simplest large-scale structure statistic on the three-point level\footnote{See \cite{Dome_2022} for a two-point level analysis in CDM, cFDM and linearised FDM.} is the third moment, also known as \textit{skewness}, which cannot be reduced to second-order statistics. The definition of PDF moments that is most widely adopted in the cosmological literature \citep{Peebles_1980, Bernardeau_2002, Szapudi_2005} is
\begin{equation}
S_p = \langle \delta^p \rangle / \langle \delta^2 \rangle^{p-1},
\end{equation}
where
\begin{equation}
\langle \delta^p \rangle = \int_{-1}^{\infty}\mathrm{d}\delta P(\delta)\delta^p,
\end{equation}
and accordingly $S_3 = \langle \delta^3 \rangle / \langle \delta^2 \rangle^2$. The overdensity distribution function $\mathrm{d}N/\mathrm{d}\delta = P(\delta)$ is defined as the normalised number of elements of the density field with a density contrast in the range $[\delta, \delta + \mathrm{d}\delta]$, and is thus related to the log PDF (which we will investigate in Sec. \ref{ss_overdens_pdfs}) via
\begin{equation}
\frac{\mathrm{d}N}{\mathrm{d}(\delta + 1)} = \frac{1}{\ln(10)}\frac{\mathrm{d}N}{(\delta + 1)\mathrm{d}\log_{10}(\delta + 1)}.
\end{equation}
Physically, $S_3$ measures the tendency of gravitational clustering to create an asymmetry between underdense and overdense regions. As clustering proceeds, there is an increased probability of having large values of $\delta$ compared to a Gaussian distribution, leading to an enhancement of the high-density tail of the PDF $P(\delta)$. As underdense regions expand and most of the volume becomes underdense, the maximum of the PDF shifts to negative values of $\delta$, and one can show \citep{Bernardeau_2002} that the maximum of the PDF to first order in the square root of the cosmic matter variance $\sigma = \sqrt{\langle \delta^2 \rangle}$ is reached at
\begin{equation}
\delta_{\mathrm{max}} \sim -\frac{S_3}{2}\sigma^2,
\end{equation}
providing useful information about the shape of the PDF.\par

We obtain error estimates on $S_3$ through jackknife resampling: The full simulation box of side length $L_{\text{box}} = 40\ h^{-1}$Mpc is divided into $4^3$ equal-sized subcubes, each time omitting one of the small cubes while calculating the statistical moments. The jackknife standard variance we adopt is
\begin{equation}
\hat{\mathrm{var}}(\hat{\theta}_{\mathrm{jack}}) = \frac{1}{N}\frac{1}{N-1}\sum_{i=1}^{N}\left(\mathrm{PV}(\mathrm{x}_{(i)})-\overline{\mathrm{PV}}\right)^2.
\end{equation}
Here, $\mathrm{x}_{(i)}$ denotes the sample but with the $i^{\mathrm{th}}$ observation removed. In our case, this translates to a subbox removal. Each \textit{pseudo-value}, $\mathrm{PV}(\mathrm{x}_{(i)})=n\hat{\theta}-(n-1)\hat{\theta}_{(i)}$, can be viewed as an estimate of $\theta = S_3$, and it is their variance that determines the jackknife standard error \citep{Efron_1982, Efron_1993}.

\section{Cosmic Web Statistics At High Redshift}
\label{s_stats}

\subsection{Mass and Volume Filling Fractions}
\label{ss_mass_vol_fractions}
\begin{figure*}
\hspace{-0.3cm}
\begin{subfigure}{0.49\textwidth}
\includegraphics[scale=0.5]{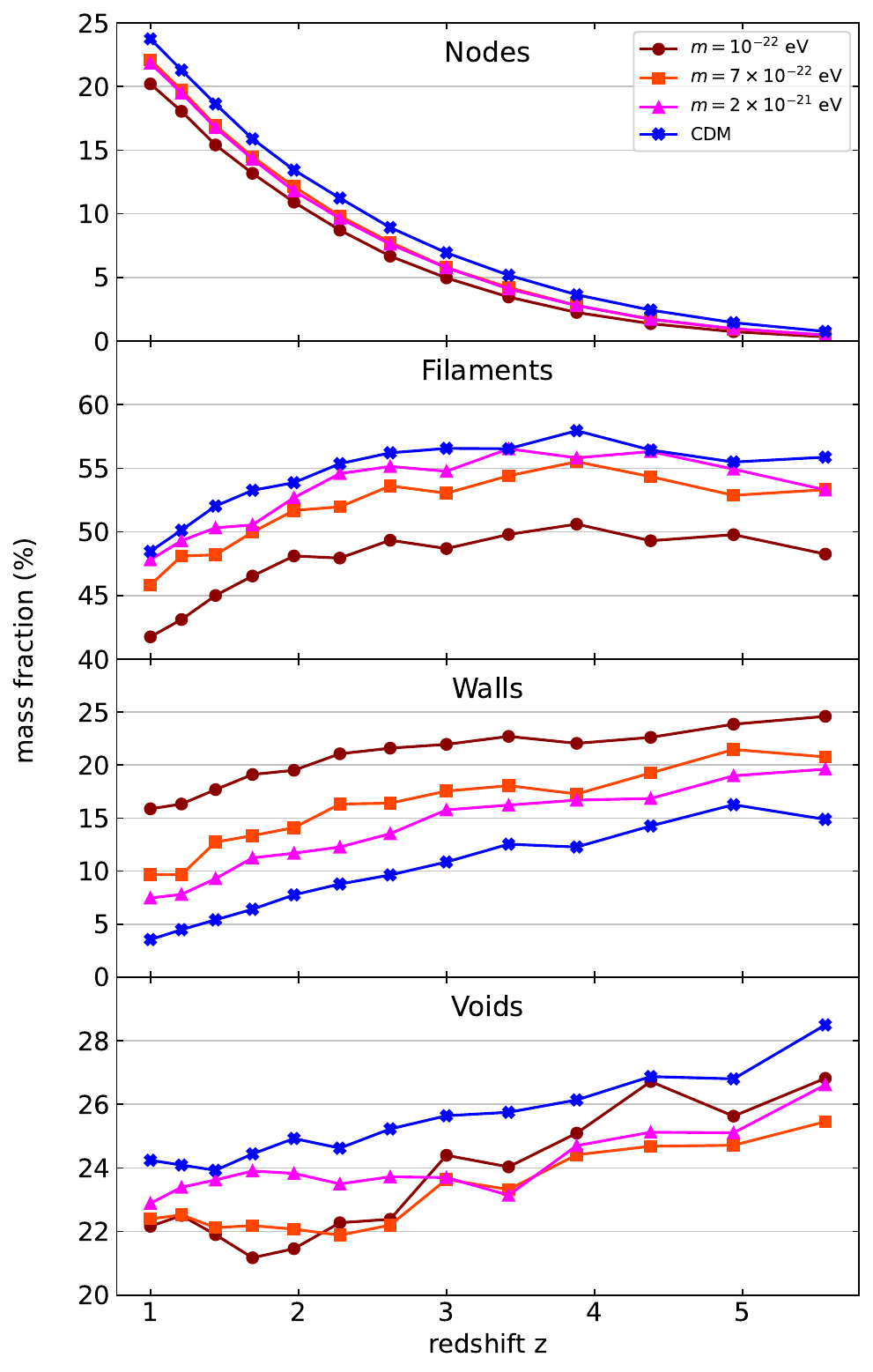}
\end{subfigure}
\hspace{0.4cm}
\begin{subfigure}{0.49\textwidth}
\includegraphics[scale=0.5]{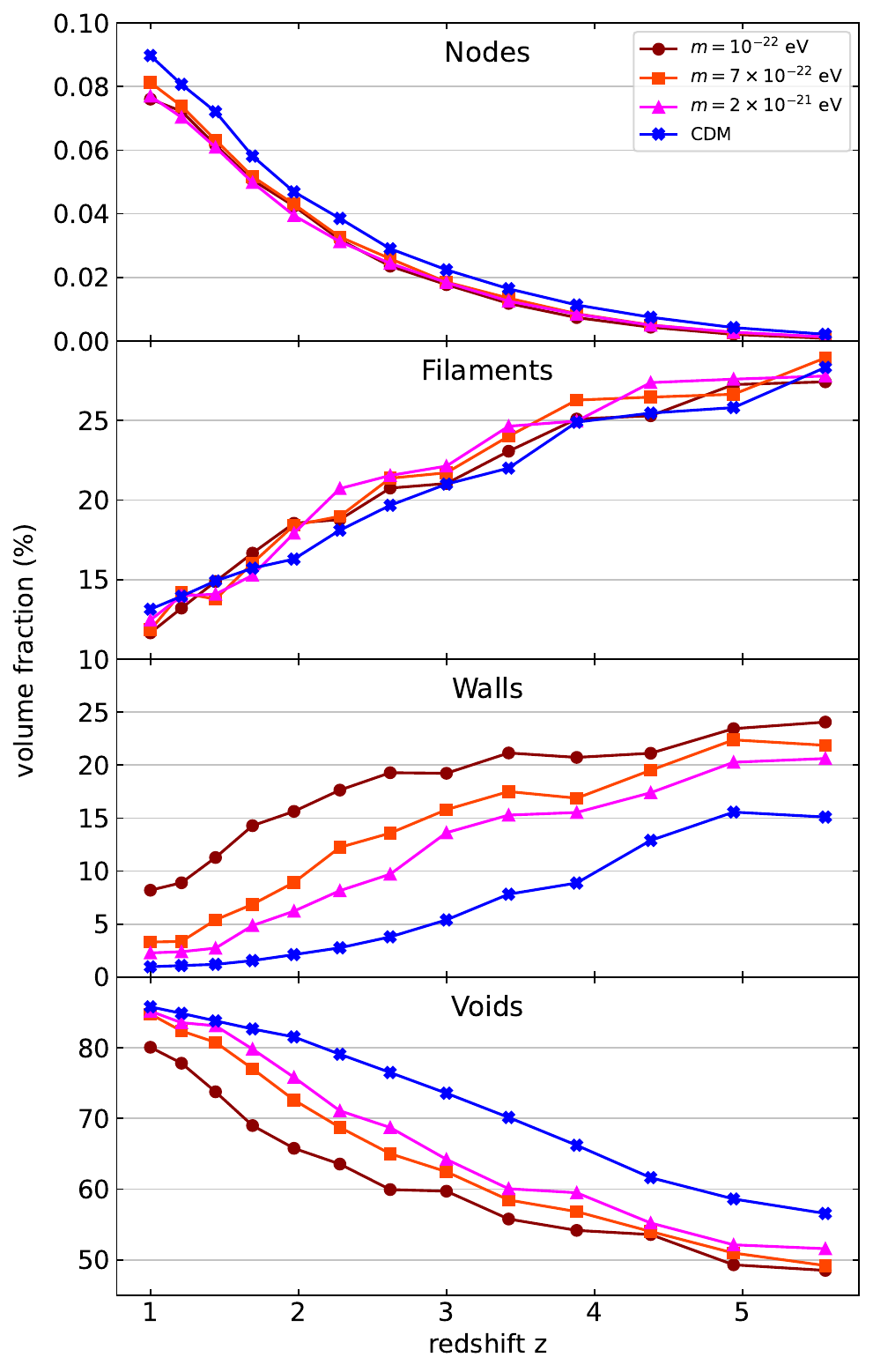}
\end{subfigure}
\caption{Evolution of the mass (left) and volume (right) filling fractions for the $N$-body CDM and cFDM runs with $1024^3$ resolution and $L_{\text{box}} = 40\ h^{-1}$Mpc. Each row represents a different NEXUS+ cosmic web environment. Cosmologies are differentiated by color as shown in the legend.}
\label{f_vol_mass_fracs}
\end{figure*}

One way of characterising the cosmic web evolution is by tracking mass and volume filling fractions of each of its components. Since each voxel is assigned only one component label, this exercise is trivial and amounts to summing up the mass or volume contained in all component voxels. We show the result in Fig. \ref{f_vol_mass_fracs} for CDM vs cFDM. The CDM results are an extension of the \cite{Cautun_2014} analysis at $z\sim 0.0-3.8$ to higher redshifts of $z\sim 1.0-5.6$ and smaller resolution scales $\Delta x$. For the overlapping redshifts $z= 1- 3.8$, we find good agreement with \cite{Cautun_2014}\footnote{The inferred CDM filament mass and volume filling fractions agree very well with \cite{Cautun_2014} despite our smaller resolution scale $\Delta x = 0.04 \ h^{-1}$Mpc. Our implementation-specific node mass fractions are systematically higher than in \cite{Cautun_2014}, e.g. at $z=3.4$, $\sim 5$\% in contrast to $\sim 0.01$\%. In \cite{Cautun_2012, Cautun_2014} the authors use a virialisation overdensity of $\Delta_{\text{vir}} = 370$ instead of Eq. \ref{e_bryan}. Also, node mass fractions are dependent on resolution scale, cf. Appendix \ref{s_res_tests}.}. To understand the evolution of the cosmic web environments in CDM, the works of \cite{Shandarin_1989, Weygaert_2008} provide good guidance: They show how matter flows out of voids towards walls, inside of which it streams towards filaments at the edges of these planar structures, which in turn channel matter towards node regions. In this simple picture which is corroborated by large-scale velocity field studies in \cite{Cautun_2014}, voids always lose mass while nodes always become more massive, establishing two opposite-trended monotonicity relations in the mass filling fractions of voids and nodes as visible in Fig. \ref{f_vol_mass_fracs}.\par

Even though walls experience both inflow and outflow of matter just as filaments, they tend to be described by decreasing mass and volume fractions at both high (Fig. \ref{f_vol_mass_fracs}) and low redshift \citep[also see][]{Cautun_2014}. Two-dimensional sheets (walls) contain $\sim 10$\% less mass and volume at $z=1$ than at $z= 5.6$ in both CDM and cFDM. By contrast, filaments tend to keep their mass filling fractions fairly constant across cosmic time until $z\sim 3$, below which they start decreasing.  Their volume fractions decrease gradually, from around $\sim 28$\% at $z= 5.6$ to around $\sim 13$\% at $z=1$. This suggests that similar mass fractions get accumulated into fewer, but more massive filaments. With the largest share of mass, cosmic filaments play a critical role in the formation of galaxies, co-determining their spin and spatial distribution \citep{Porter_2005, Malavasi_2020, Poudel_2017}. It stresses the importance of revisiting standard theories which assume that halo environments in which protogalaxies form play the dominant role in shaping the properties of galaxies \citep{Wilman_2012, Behroozi_2010}.\par 

How DM is distributed across different components of the cosmic web depends on the DM model at hand and thus the cosmology. Here we find that the lower the axion mass $m$, the lower the relative share of mass in nodes\footnote{At first approximation, it is the decrease in the number of small nodes that reduces cFDM node mass fractions, in analogy to cFDM halo mass functions that are also suppressed at the small-mass end (cf. Sec. \ref{ss_halo_distros}).} and filaments and the higher the mass share in walls when compared to CDM. To be precise, the mass filling fraction of filaments exhibits a $\sim 6-8$\% decrease between CDM and the $m=10^{-22}$ eV model across most of the redshifts investigated. This is mainly compensated for by a $\sim 8-12$\% increase in mass filling fractions of walls between CDM and the $m=10^{-22}$ eV model. The redistribution of DM to higher-dimensional structures, i.e. sheets, is a manifestation of the loss of small-scale power in the primordial and also evolved DM distributions.\par

The DM budget in each cosmic web environment also affects large-scale tidal forces in the Universe, shaping the evolution of halos and galaxies. In the same way that tidal torque theory predicts quadrupolar patterns in the vorticity field around the saddle points of cosmic filaments \citep{Codis_2015}, tidal forces give rise to dipolar patterns around cosmic sheets. Consequently, such dipolar features are expected to be more pronounced in cFDM cosmologies than in CDM. Sheet-like morphologies are attributed increased importance in the gas dynamics of FDM as well, since massive gas pancakes are predicted to be the sites of first star formation \citep{Kulkarni_2022}. Note, however, that global mass filling fractions are anisotropy-agnostic and by construction gloss over the enhanced matter distribution anisotropies apparent in cFDM cosmologies \citep{Dome_2022}. Reliable tidal force predictions would necessitate an analysis of anisotropic geometries and are thus beyond the scope of this paper.\par 
 
\subsection{Overdensity PDFs}
\label{ss_overdens_pdfs}
\begin{figure}
\hspace{-0.2cm}
\includegraphics[scale=0.53]{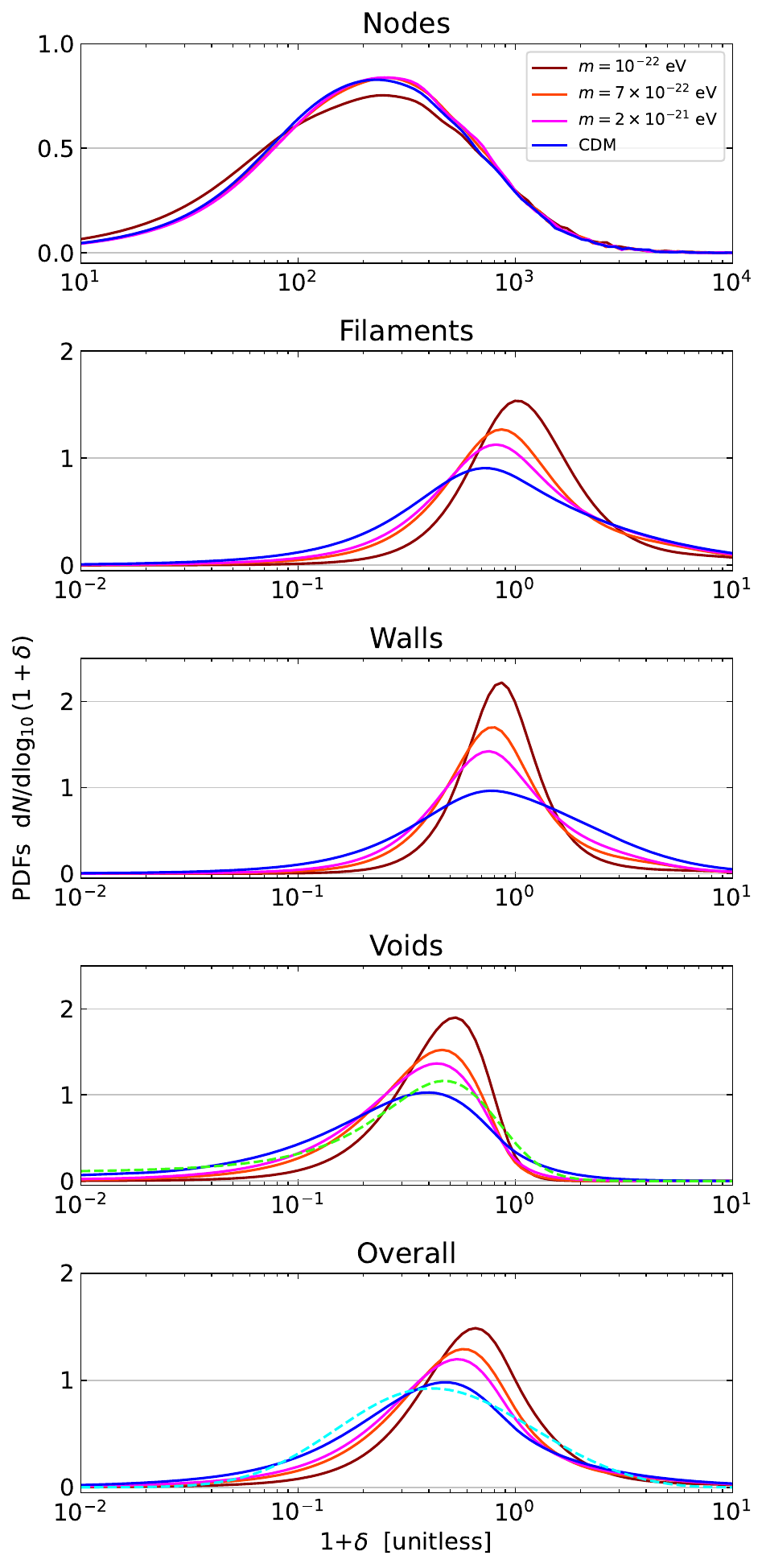}
\caption{Log overdensity PDFs for the $N$-body CDM and cFDM runs with $1024^3$ resolution and $L_{\text{box}} = 40\ h^{-1}$Mpc at redshift $z=3.9$. Cosmologies are differentiated by color as shown in the legend. The first four rows represent different NEXUS+ cosmic web environments while the last row shows the overall log overdensity PDFs. The dashed green curve (fourth row) is the CDM best-fit result using the \protect\cite{Miralda_2000} fitting formula for the void log overdensity PDF while the dashed cyan curve (bottom row) is the CDM best fit among the family of reversed Weibull distributions \protect\citep[cf.][]{Repp_2018}.}
\label{f_dens_pdfs}
\end{figure}
The simplest way of characterising the variation of the matter content across environments is via density distributions. As in the rest of this work, we use the CIC density to obtain the probability density function (PDF) of the log overdensity field $\log_{10}(1+\delta)$. In Fig. \ref{f_dens_pdfs}, the PDF is segmented into morphological components. Let us again start the discussion with CDM. We find that various cosmic environments are characterised by different values of the log overdensity field. Node regions have by far the highest PDF median at around $1+\delta \sim 300$. Filaments also predominantly represent overdense environments as can already be predicted within the Zel'dovich formalism \citep{Pogosyan_1998} and has been found by e.g. \cite{Aragon_2010}. Walls and especially voids are more likely to be found in underdense environments. The large widths of the distributions give rise to significant overlaps between the log overdensity PDFs of different components. A simple density threshold \citep{Shandarin_2004, Dolag_2006} is thus only sufficient to identify cosmic nodes but cannot be used to differentiate between the remaining components.\par 

To theorize the void PDF, one can make the simple approximation \citep{Miralda_2000} that matter expands in voids at a constant velocity. In the absence of tidal forces, this approximation holds. For Gaussian initial conditions, the overdensity distribution is thus $P_V(\Delta) \propto \exp(-C\Delta^{-4/3})\Delta^{-8/3}$, where $\Delta = 1+\delta$. This distribution can also be obtained as the $\Delta \ll 1$ limit of the \cite{Miralda_2000} fitting formula 
\begin{equation}
\label{e_miralda}
P_V(\Delta) = A \exp\left[-\frac{(\Delta^{-2/3}-C_0)^2}{2(2\delta_0/3)^2}\right]\Delta^{-\beta}.
\end{equation}
In Fig. \ref{f_dens_pdfs}, we use parametrization \eqref{e_miralda} and add the best-fit result for CDM. We find rather poor agreement, which could hint at the presence of tidal forces but also imperfect segmentation into nodes, filaments, walls and voids. In the NEXUS+ formalism, voids are simply the complement to nodes, filaments and walls, rather than being extracted directly.\par 

The overall log overdensity PDF is shown in the last row of Fig. \ref{f_dens_pdfs} and is amenable to a certain level of (semi-)analytical scrutiny. In fact, there are various attempts \citep{Klypin_2018, Repp_2018, Uhlemann_2020} at predicting the distribution of the cosmological density field for CDM and extracting the cosmological information stored in the (log) density field. \cite{Repp_2018} lay out the necessary ingredients to construct the predicted log overdensity PDFs, hence we will focus on their generalised extreme value (GEV) model that they built for scale-free cosmologies.\par

A subclass of the GEV distributions, the reversed Weibull distribution assumes the form
\begin{equation}
P(\log_{10} \Delta) = \frac{1}{\sigma_{\text{GEV}}}t(\log_{10} \Delta)^{1+\xi}e^{-t(\log_{10} \Delta)},
\end{equation}
where $t(\log_{10} \Delta) = \left(1+\frac{\log_{10} \Delta-\mu_{\text{GEV}}}{\sigma_{\text{GEV}}}\right)^{-1/\xi}$ and $\xi < 0$. Combined with fits for the mean, variance and skewness of the log overdensity field $\log_{10} \Delta$, \cite{Repp_2018} show that at lower redshifts of $z\lesssim 2$ and smoothing scales down to $2 \ h^{-1}$Mpc, the parametrised reversed Weibull distribution provides an excellent fit to the Millennium Simulation (MS) results \citep{Springel_2005}. However, it is a very poor fit (not shown) at our redshifts of interest $z\sim 1.0-5.6$ and resolution scale $L_{\text{box}}/N_{\text{lin}} = 0.04 \ h^{-1}$Mpc. To make matters worse, the whole \textit{family} of reversed Weibull distributions is inadequate at our redshifts and smoothing scale, judging by the poor best-fit\footnote{The best-fit curve in the family of reversed Weibull distributions (dashed cyan in Fig. \ref{f_dens_pdfs}) yields a reduced chi square $\chi^2_{\nu} = \chi^2/\nu = 348.9 \gg 1$ with $\nu = N - 3$ where $N = 1024^3$ is the number of samples, indicating a very poor fit. The hypothesis that the data are from a reversed Weibull population should thus be rejected. Note that the log overdensity samples are correlated and thus do not constitute independent realisations of an underlying distribution. The Pearson chi square goodness of fit is meant as an illustration only.} for the CDM log overdensity PDF shown in Fig. \ref{f_dens_pdfs}. We conclude that the best analytical models to date for log overdensity PDFs need further improvement to provide a reliable model at redshifts $z\gtrsim 2$ and smoothing scales smaller than $2 \ h^{-1}$Mpc, let alone for cosmologies that break the hierarchical nature of the cosmic web such as the cFDM models. Since such an extension is beyond the scope of this paper, we postpone it to future work and contend with showing the resolution scale dependence of log overdensity PDFs in \mbox{Appendix \ref{s_res_tests}}.\par 
\begin{figure}
\vspace{0.02cm}
\hspace{-0.47cm}
\includegraphics[scale=0.52]{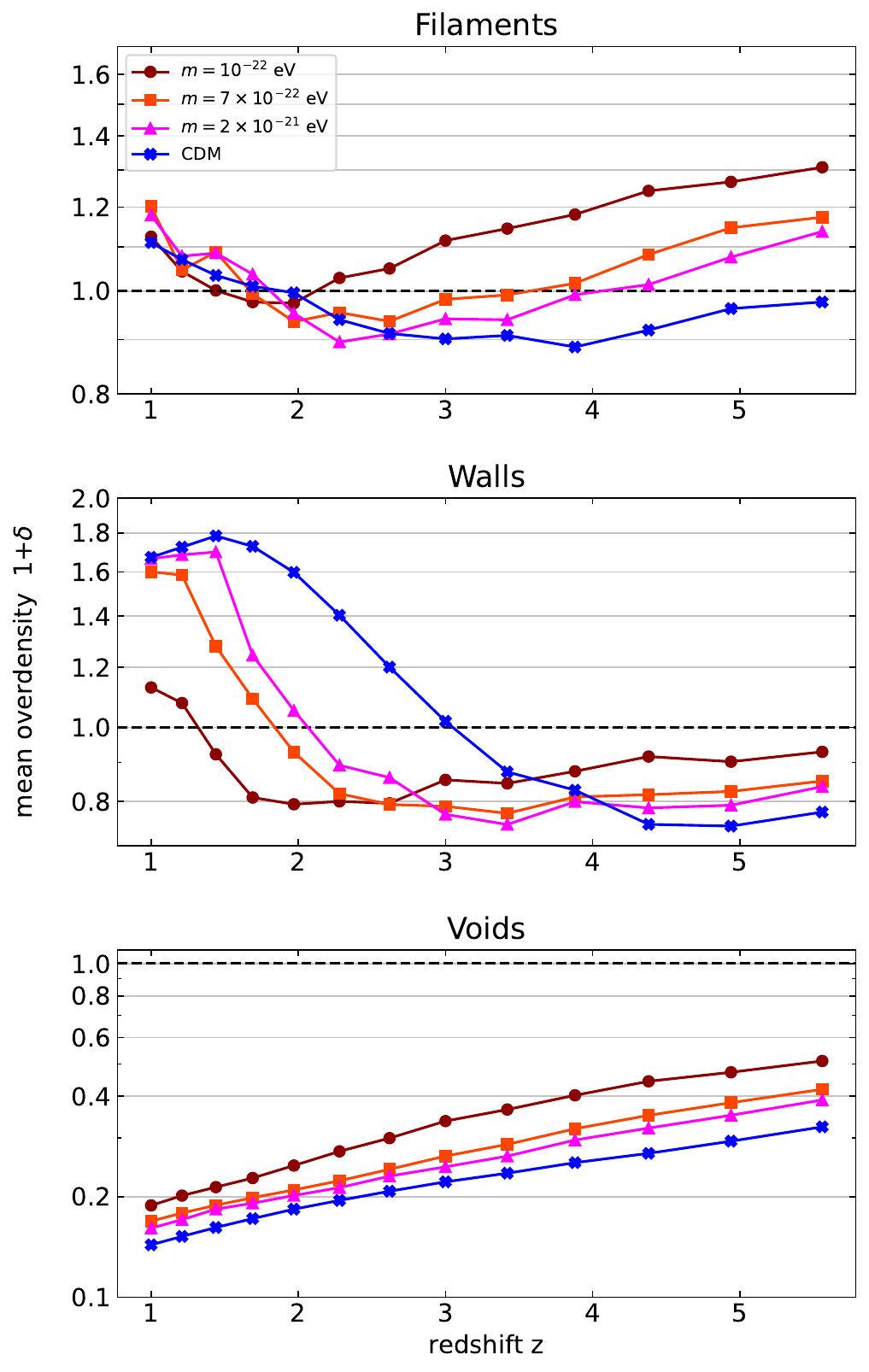}
\caption{Median of log overdensity PDFs for the $N$-body CDM and cFDM runs with $1024^3$ resolution and $L_{\text{box}} = 40\ h^{-1}$Mpc across redshifts $z\sim 1.0-5.6$. Cosmologies are differentiated by color as shown in the legend. Results are shown for three NEXUS+ morphological components: filaments (first row), walls (second row) and voids (third row).}
\label{f_dens_pdfs_medians}
\end{figure}
The cFDM models show significantly narrower distributions around the median values than in CDM, except for node environment PDFs which are fairly insensitive to a primordial power spectrum cutoff. Filament environment PDFs, for instance, have their full width at half maximum (FWHM) decrease from $0.94$ dex for CDM to $0.50$ dex for $m=10^{-22}$ eV cFDM. For the overall PDFs (last row in Fig. \ref{f_dens_pdfs}), the corresponding numbers read $0.84$ dex vs $0.55$ dex. Intuitively, this can be explained as follows: In the case of filaments, small-scale structure is typically associated with tenuous tendril-like features or substructure at the periphery of more major filaments, which get washed out as the axion mass $m$ gets reduced. The suppression of the high-overdensity tail results from the delayed formation of large-scale structure and high-mass halos in particular compared to CDM \citep[see e.g.][]{Safarzadeh_2018}. This effect is most striking for walls which in cFDM have a higher share of mass, see Fig. \ref{f_vol_mass_fracs}. With suppression at both ends, the PDF is more narrow. For all environments except nodes, the narrower distribution with a strong mid-range peak illustrates that density minima are more shallow. In the case of voids, this is a well-known result \citep{Yang_2015} that is independent of the adopted cosmic web dissection algorithm.\par

To quantify the dependence of log overdensity means on cosmology over a range of redshifts, in Fig. \ref{f_dens_pdfs_medians} we plot the mean of the log overdensity PDFs for filaments, walls and voids. We refrain from showing node means. Beyond $z\sim 3$, node means attain values $1+\delta \sim 300$ and due to the near-constancy of the virial collapse threshold $\Delta_{\text{vir}}$ (cf. Eq. \eqref{e_bryan}) above $z\sim 3$, they exhibit little variation with redshift and cosmology. Below $z\sim 3$, node means grow in tandem with $\Delta_{\text{vir}}$. For filaments, walls and voids, the shifting of the mean to higher overdensities with decreasing axion mass $m$ is very pronounced, with increases up to $\sim 55$\% in the case of voids. Void overdensity means decrease as the Universe evolves, in agreement with the simplified picture of voids becoming more and more devoid of matter as demonstrated in Fig. \ref{f_vol_mass_fracs} \citep[see also][]{Haarlem_1993, Weygaert_2008}. On the other hand, filament and wall overdensity means drop less quickly with decreasing $z$ and even increase below a cosmology- and resolution scale-dependent redshift. For instance, the CDM filament mean starts increasing below $z\sim 3$, a trend that grows more pronounced towards lower redshift \citep[also see][]{Cautun_2014}. At high redshift, walls peak below the mean background density of the Universe as already predicted in the Zel'dovich framework \citep{Pogosyan_1998}. However, the means cross the $\delta = 0$ line at lower redshift, namely at $z=3$ for CDM and $z=1.3$ for the rather extreme cFDM model with $m=10^{-22}$ eV. The growth of the filament and wall means with cosmic time at lower redshift is a result of the mass in the filamentary and wall networks getting concentrated in fewer but more massive structures.\par

It is known that the distribution of log (over-)densities provides powerful constraints on the matter density $\Omega_m$, the clustering amplitude $\sigma_8$ and the total neutrino mass $M_{\nu}$ \citep{Uhlemann_2020}. One of the reasons for their success is that the log transformation effectively erases non-linear evolution from the power spectrum \citep{Neyrinck_2009}, despite $\log_{10}(1+\delta)$ being only approximately Gaussian. We now find that there is a strong dependence of log overdensity medians and shapes on DM models too, which suggests that the distribution of log overdensities constitutes a promising new testbed for constraining DM scenarios. Clearly, the challenge lies in estimating the bias of the log power spectrum extracted from discrete surveys based on tracers of the DM field, typically galaxies. However, there has been considerable progress in the construction of such estimators \citep{Repp_2018_bias, Repp_2019}.\par
\begin{figure}
\vspace{0.24cm}
\hspace{-0.4cm}
\includegraphics[scale=0.51]{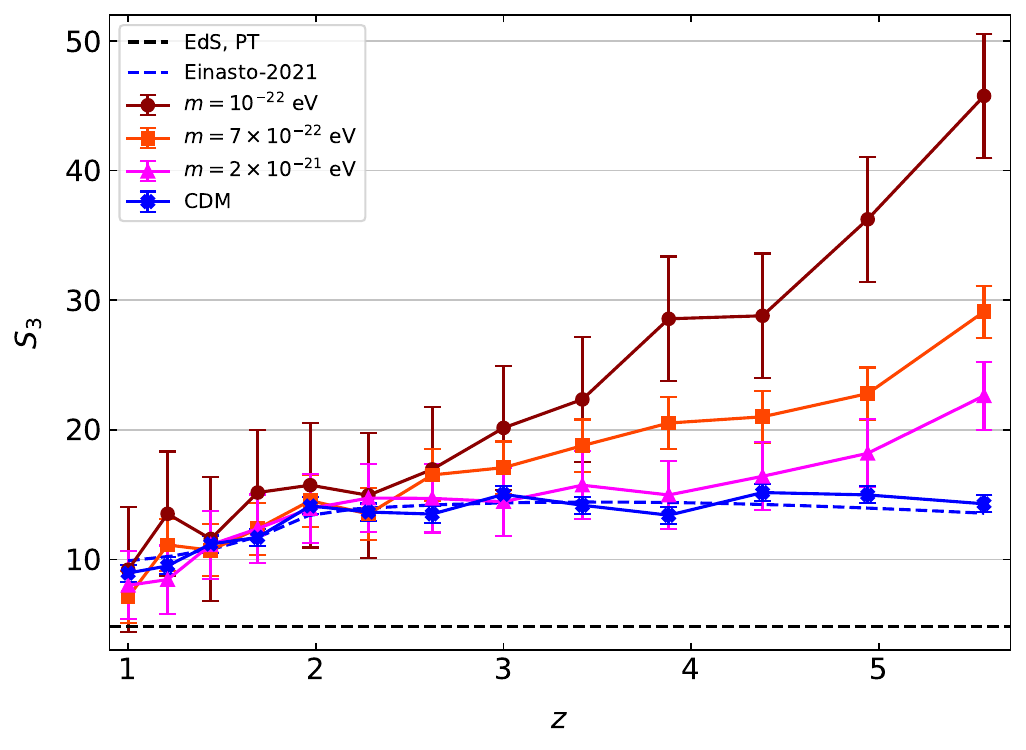}
\caption{Skewness $S_3$ of overdensity PDFs for the $N$-body CDM and cFDM runs with $1024^3$ resolution and $L_{\text{box}} = 40\ h^{-1}$Mpc across redshifts $z\sim 1.0-5.6$. Cosmologies are differentiated by color as shown in the legend. There is no conditioning on cosmic environment. Error estimates from jackknife resampling with $4^3$ subboxes are marked. The prediction $S_3 = 34/7$ by \protect\cite{Peebles_1980}, based on linear PT, for the Einstein-de Sitter (EdS) model (for which $\Omega_m = 1.0$) is shown for comparison.}
\label{f_dens_pdfs_skewness}
\end{figure}
In order to gain insight into the asymmetries of the overdensity distribution function, we briefly focus on the third moment of the unconditioned PDF $P(\delta)$, also known as skewness (see Sec. \ref{ss_skewness}). We focus on the overall skewness $S_3$ as is common, without conditioning on environment. The evolution of both the CDM and cFDM model Universes starts off from a Gaussian random field that is symmetrical around the mean density, that is, positive and negative deviations from the mean density are equally probable (hence $S_3=0$). Gradually, the overdensity field $\delta$ becomes asymmetric. It appears as soon as nonlinearities start to play a role because, in non-linear large-scale structure theory, underdense regions evolve less rapidly than overdense regions \citep{Bernardeau_2002}.\par 

In Fig. \ref{f_dens_pdfs_skewness}, we present skewness estimates for CDM and cFDM cosmologies (the latter for the first time). We find that our CDM $S_3$ estimates are well traced by the fitting formula devised by \cite{Einasto_2021} for the fiducial CDM cosmology. For a fixed smoothing scale $R$ (in our case $R = \Delta x = L_{\text{box}}/N_{\text{lin}} = 0.04 \ h^{-1}$Mpc) and in the redshift range $z\sim 2.5-5.0$, $S_3$ finds itself close to the plateau regime of its \textit{evolutionary track}. At both higher redshift $z\gtrsim 5.0$ and lower redshift $z\lesssim 2.5$, $S_3$ assumes lower values and eventually flattens off below $z\sim 1.0$. Using $N$-body simulations, \cite{Einasto_2021} show that $S_3$ is not merely a function of the square root of the cosmic matter variance, $\sigma$, but also explicitly dependent on either $z$ or the smoothing scale $R$. Since two quantities in the set $\lbrace \sigma, z, R\rbrace$ determine the third one, the dependence of $S_3$ can be written either way. Models expressed solely as a function of $\sigma$ (such as the lognormal distribution) thus cannot possibly account for the evolution of the PDF with redshift.\par 

For cFDM, we find that the $S_3$ estimates are systematically higher than the CDM ones, especially at higher redshift. At $z=5.6$, the fractional difference\footnote{Error bars are higher for cFDM models since there are more CIC cells that have zero density, which cannot be properly captured by the PDF $P(\delta)$ (also discussed in \cite{Einasto_2021}).} in $S_3$ between $m=10^{-22}$ eV cFDM and CDM is $(S_3^{10^{-22}\ \mathrm{eV}}-S_3^{\mathrm{CDM}})/S_3^{\mathrm{CDM}} = 2.20 \pm 0.35$, which is different from zero at a level of $\sim 6 \sigma$.\par 

The fact that $S_3$ is lower for power spectra with more small-scale fluctuations (CDM) than those with fewer small-scale fluctuations (cFDM) has already been theorised by \cite{Bernardeau_2002} using the following argument: Dating back to earlier works \citep{Bernardeau_1994, Bernardeau_1995}, it has been noted that the dependence of skewness with the shape of the power spectrum comes from a mapping between Lagrangian space, in which the initial size of the perturbation is determined, and Eulerian space. For a given filtering scale $R$, overdense regions with $\delta > 0$ come from the collapse of regions that had initially a larger size while underdense regions with $\delta < 0$ come from initially smaller regions. In cFDM that has a lack of small coherent regions in the primordial density field, the asymmetry between under- and overdense regions in the evolved density field is greater than in CDM. In Fig. \ref{f_dens_pdfs_skewness}, this effect on the skewness $S_3$ is quantified. In cFDM, non-linear structure formation \citep[e.g. halo mass assembly,][]{Khimey_2021} proceeds faster than in CDM. In the range $z\sim 2.5-5.0$ where the CDM evolutionary track is around its plateau for the chosen smoothing length $R$, cFDM is thus already past its plateau and $S_3$ decreases monotonically with cosmic time before flattening off at $z\sim 1.0$.\par 

\subsection{Halo Mass Distributions}
\label{ss_halo_distros}

According to standard theories of structure formation, DM halos play a crucial role in galaxy formation. However, as we alluded to in Sec. \ref{ss_mass_vol_fractions}, cosmic environments co-determine the formation of galaxies not least because of large-scale tidal forces. For instance, the enhancement in clustering induced by correlations between halo assembly history and large-scale environment at fixed halo mass is readily observed in cosmological simulations yet difficult to detect in observations \citep[halo \& galaxy assembly bias]{Sunayama_2022, Xu_2021}. Here, we investigate the differences in the halo population across the cosmic web components which in turn are suggestive of variations with large-scale environment in the population of galaxies and their properties.\par

\begin{figure}
\hspace{-0.5cm}
\includegraphics[scale=0.52]{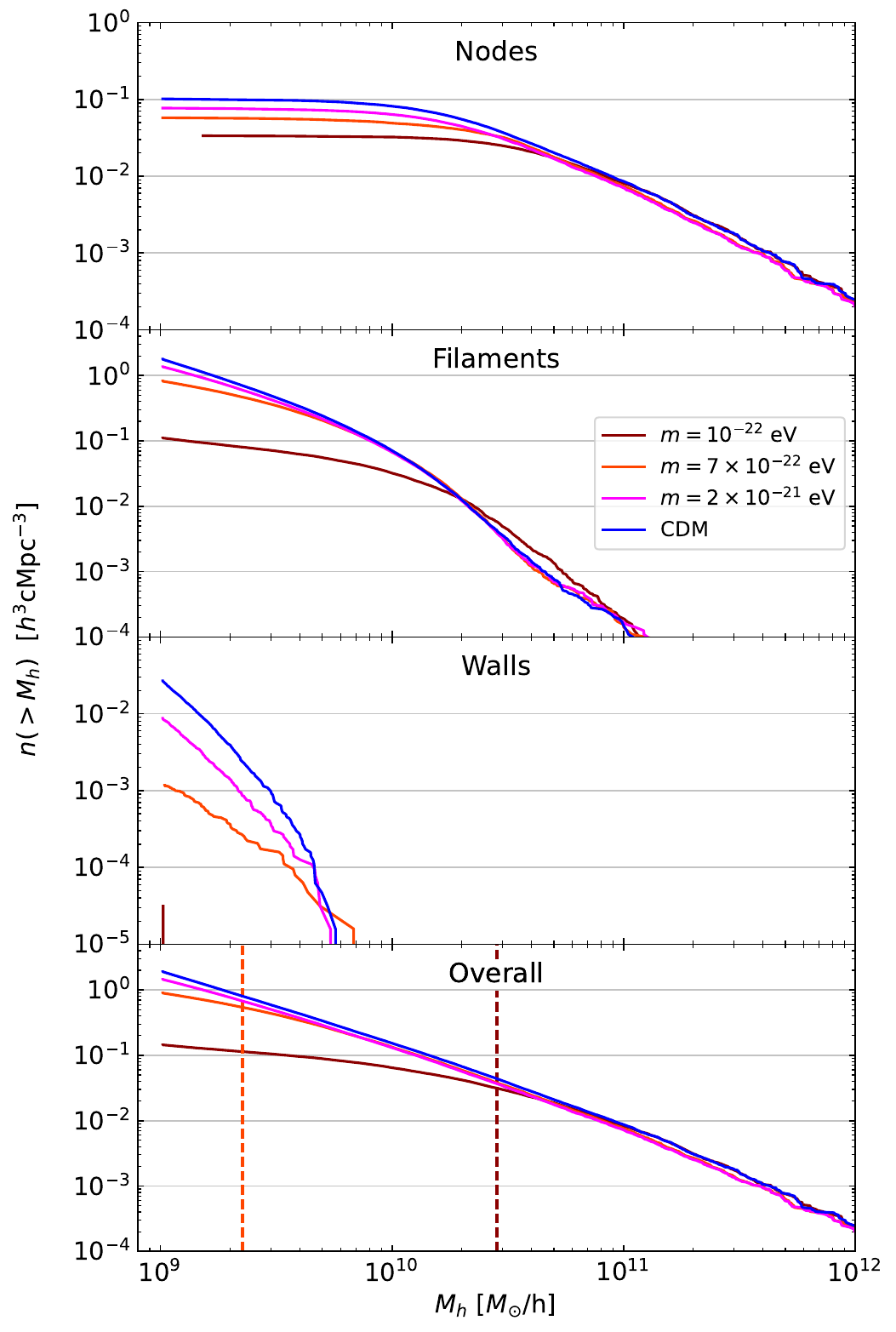}
\caption{Cumulative halo mass functions (cHMFs) for the $N$-body CDM and cFDM runs with $1024^3$ resolution and $L_{\text{box}} = 40\ h^{-1}$Mpc at redshift $z=3.9$, split according to the NEXUS+ environment in which the halo resides, indicated for each panel. The last row shows the overall cHMFs; vertical dashed lines denote the half-mode mass $M_{1/2}$ \citep{Marsh_2016} of the $m=10^{-22}$ eV and the $m=7\times 10^{-22}$ eV models (for $m=2\times 10^{-21}$ eV cFDM, $M_{1/2} = 5.77 \times 10^8 \ h^{-1} M_{\odot}$ is off-scale).}
\label{f_hmfs}
\end{figure}

Fig. \ref{f_hmfs} shows the cumulative halo mass function (cHMF) segmented into cosmic web environments, at $z=3.9$. We find that the most massive halos across all cosmologies are exclusively found in node regions, especially beyond $M_h \sim 10^{11} \ h^{-1} M_{\odot}$. The vast majority of halos that are not located in nodes are filament halos, which start to dominate the cHMF below about $M_h\sim 2\times 10^{10}\ h^{-1} M_{\odot}$. Halos in walls and voids (not shown) represent a substantial share of the halo population only at the lowest resolved masses below $M_h \sim 2\times 10^{9} \ h^{-1} M_{\odot}$. In particular, since this behavior is exhibited regardless of cosmology it implies that very few luminous galaxies and quasars are and will be observed in cosmic sheets with current and upcoming galaxy/QSO redshift surveys such as SDSS SEQUELS \citep{Myers_2015}, the DESI Bright Galaxy Survey \citep[BGS]{Zarrouk_2021} and JWST Advanced Deep Extragalactic Survey \citep[JADES]{Linzer_2020}.\par

In analogy to WDM \citep{Schneider_2012} and bona-fide FDM HMF analyses \citep{May_2022}, we confirm that cFDM cosmologies have fewer small-mass halos compared to CDM but here we quantify the environment-conditioned cHMFs. All node-, filament- and wall-conditioned cHMFs exhibit a strong suppression in cFDM cosmologies, but for some environments this occurs well above the half-mode mass $M_{1/2}$ \citep{Marsh_2016}. As seen in Fig. \ref{f_hmfs} at $z=3.9$, the node-conditioned cHMF of the $m=7\times 10^{-22}$ eV model exhibits a $50$\% reduction below a mass of $M_h \sim 10^{10} \ h^{-1} M_{\odot}$ while the half-mode mass is $M_{1/2} = 2.3\times 10^{9} \ h^{-1} M_{\odot}$. For wall halos in $m=7\times 10^{-22}$ eV cFDM, we observe a $>50$\% reduction below a mass of $M_h \sim 4 \times 10^{9} \ h^{-1} M_{\odot}$. Thus, if the given environment is not dominant on the mass scale considered, the cFDM suppression can turn out stronger than naively expected from $M_{1/2}$.\par 

We also observe that for walls (subdominant environment), the cFDM suppression of the conditioned cHMFs is stronger than for nodes and filaments. At the smallest resolved halo mass of $M_{\text{min}} \sim 10^{9} \ h^{-1} M_{\odot}$, compared to CDM the $m=7\times 10^{-22}$ eV cFDM model exhibits a $\sim 1.1$ dex suppression in the wall-conditioned cHMF. For both the node-conditioned and filament-conditioned HMF, however, the corresponding suppression is less than $\sim 0.5$ dex. In simple terms, walls and voids (not shown) are disproportionately more devoid of halos in cFDM cosmologies than in CDM. In addition, the filament-conditioned cHMF of the rather extreme $m=10^{-22}$ eV model features a slight enhancement above $M_h \sim 2 \times 10^{10} \ h^{-1} M_{\odot}$, though due to the smallness of this effect we refrain from attributing physical significance to it.

\section{Conclusions}
\label{s_conclusions}
In this work, we present a cosmic web analysis for a cosmology with a small-scale suppression of power such as FDM or WDM for the first time. We compare the cosmic web structure in CDM vs three instances of cFDM with axion masses $m=10^{-22}, \ 7\times 10^{-22}, \ 2\times 10^{-21}$ eV by analysing a suite of cosmological $N$-body simulations. For the analysis, we use our independent implementation of the NEXUS+ segmentation algorithm \citep{Cautun_2012}. We present the overall mass and volume filling fraction of cosmic web environments, their density distributions and conditioned halo mass functions. We summarise our main conclusions as follows:  

\begin{enumerate}[label={(\alph*)}]
\item We recover in cFDM some general trends of CDM such as fairly time-independent mass content in progressively fewer cosmic filaments. In addition, we observe that while lower-dimensional structures such as nodes and filaments contain less mass in cFDM cosmologies, the mass filling fraction of cosmic sheets is inversely proportional to the adopted axion mass $m$. In cFDM, it is thus expected that the tidal force-induced dipolar patterns in the vorticity field \citep{Codis_2015} around cosmic sheets are more pronounced than in CDM, affecting the angular momentum properties of halos and galaxies. For the $m=7 \times 10^{-22}$ eV model and across most of the investigated redshift range of $z\sim 1.0-5.6$, we observe a $\sim 8-12$\% increase in the mass filling fraction of walls compared to CDM, compensate for mainly by a $\sim 6-8$\% reduction in the mass filling fraction of filaments.
\item Given a cosmological model, low-order moments of $P(\delta)$ on small smoothing scales (we adopt $R =0.04 \ h^{-1}$Mpc) can be approximated only poorly using perturbation theory, let alone the full information contained in the PDF $P(\delta)$. However, numerical $N$-body simulations provide powerful insights into $P(\delta)$ for both CDM \citep[e.g.][]{Einasto_2021} and cFDM (this work). All morphological components except virialisation condition-based cosmic nodes feature more concentrated log overdensity PDFs $P(\log_{10}(1+\delta))$ in cFDM cosmologies, with strong mid-range peaks. At high redshift, the PDF mean (first-order moment) is systematically higher than in CDM. This is a result of the chipping off of the low overdensity tail as $m$ is reduced, which in turn is a reflection of the lack of small-scale structure in FDM-like cosmologies.\par 
\item Skewness estimates $S_3$ (rescaled third-order moment) of the unconditioned PDF $P(\delta)$ in cFDM are systematically higher than in CDM, especially at the highest investigated redshift $z\sim 5.6$ where $m=10^{-22}$ eV cFDM differs from CDM by $\sim 6 \sigma$. The (over-)density PDF provides powerful constraints on the matter density $\Omega_m$, the clustering amplitude $\sigma_8$ and the total neutrino mass $M_{\nu}$ \citep{Uhlemann_2020}, while new methodologies are underway to estimate the bias of the log power spectrum for discrete surveys \citep{Repp_2018_bias}. Due to their additional sensitivity to small-scale cutoffs in the primordial power spectrum, we thus recommend (log) matter PDFs as a new testbed for constraining FDM and, potentially, other alternative DM models such as WDM. Due to its similarities to cFDM, our results already shed light on structure formation in WDM cosmologies, for which (log) density PDFs and their skewness remain to be investigated.
\item Cosmic filaments have the highest share of mass even for the rather extreme $m=10^{-22}$ eV model across the $z\sim 1.0-5.6$ range investigated, which is also reflected in environment-conditioned cHMFs. At $z=3.9$, for instance, we find that filament halos start to dominate the cHMF below about $M_h\sim 2\times 10^{10}\ h^{-1} M_{\odot}$. While the suppression of small-scale power naturally leads to a suppression in the cHMFs \citep{Schneider_2012, May_2022}, here we quantify said suppression in the environment-conditioned cHMFs and find that it can occur well above the half-mode mass $M_{1/2}$. E.g. for node halos in $m=7\times 10^{-22}$ eV cFDM, we observe a $\sim 50$\% reduction below a mass of $M_h \sim 10^{10} \ h^{-1} M_{\odot}$ while the half-mode mass is $M_{1/2} = 2.3\times 10^{9} \ h^{-1} M_{\odot}$. Importantly, for halos in walls and voids the suppression is stronger than for those in nodes and filaments (see Fig. \ref{f_hmfs}). In other words, walls and voids that already host few halos in CDM are disproportionately more devoid of halos in cFDM cosmologies.
\end{enumerate}

Any exponential-like cutoff in the primordial power spectrum at the small-scale end is heavily constrained observationally, cf. \cite{Dome_2022} for an overview of recent and forecast astrophysical constraints. However, a more reliable verdict on DM scenarios with small-scale cutoffs and mixed DM models is yet to be made. \href{https://webb.nasa.gov/}{JWST} and in particular the JADES survey \citep{Linzer_2020} will revolutionise our understanding of high-redshift galaxies, and their ages, correlation functions as well as UV luminosity functions will help improve DM constraints upon comparison with predictions from simulations \citep[e.g.][]{Esmerian_2021}. This will come in handy as the search for a primordial small-scale cutoff is especially promising in the more pristine and quasi-linear high-redshift cosmic web. What we seek to provide here is a thorough understanding of how FDM-like cosmologies impact the high-$z$ cosmic web, which is necessary to improve the reliability of high-$z$ DM constraints.\par 

The next-generation of line intensity mapping surveys \citep[LIM,][]{Kovetz_2017} and 21cm experiments \citep{Trott_2019} will complement JWST and make use of hydrogen (HI) and other atoms as tracers of the cosmic web. In the post-reionisation era, while being too faint for most available telescopes, the small amount of gas and dust in the faint galaxies tracing filamentary structures of the cosmic web will be made visible upon integrating their cumulative emission \citep{Fonseca_2019}. During the era of reionisation and before, the order-of-magnitude improvement in survey depth that the \href{https://www.skao.int/}{SKA} will bring compared to existing interferometers will allow to probe column densities of $N_{\text{HI}}=10^{18} \ \text{cm}^{-2}$ and below over large areas on the sky at sub-arcminute resolution, advancing our understanding of the distribution of neutral hydrogen in the IGM and thus the cosmic web \citep{Popping_2015, Kale_2016}. These instruments will provide novel tests for the newly emerging picture in which cosmic filaments play an instrumental role in shaping galaxy formation and evolution \citep{Malavasi_2016, Mandelker_2018, Wang_2021}, making the cosmic web bound to be subject to ever more scientific scrutiny.\par 

\section{Acknowledgements}
We are grateful to Krishna Naidoo for very helpful feedback on our manuscript. It is a pleasure to thank Debora Sijacki for enriching conversations. We are also grateful to Sophie Koudmani and Martin Bourne for providing help on the ins and outs of \scshape{Arepo}\normalfont . TD acknowledges support from the Isaac Newton Studentship and the Science and Technology Facilities Council under grant number ST/V50659X/1. AF is supported by the Royal Society University Research Fellowship. NS gratefully acknowledges the support of the Research Foundation - Flanders (FWO Vlaanderen), grant 1290123N. The simulations were performed under DiRAC project number ACSP253 using the Cambridge Service for Data Driven Discovery (CSD3), part of which is operated by the University of Cambridge Research Computing on behalf of the \href{https://dirac.ac.uk}{STFC DiRAC HPC Facility}. The DiRAC component of CSD3 was funded by BEIS capital funding via STFC capital grants ST/P002307/1 and ST/R002452/1 and STFC operations grant ST/R00689X/1. DiRAC is part of the National e-Infrastructure.

\section{Data Availability}
\label{s_data_availability}
High-level data products are available upon reasonable request.

\bibliographystyle{mnras}
\bibliography{refs}

\appendix
\section{Quasi-Virialisation of Filaments}
\label{s_vir}
An obvious question to ask is how to define the extents of a cosmic filament in simulations and observations. For halos, the virialisation condition is typically estimated using spherical collapse theory by extrapolating the linear theory solution to beyond shell-crossing. With respect to the mean overdensity of the universe, it is given by \citep{Bryan_1998}
\begin{equation}
\Delta_{\text{vir}} = \frac{18\pi^2+82x-39x^2}{\Omega_0}-1,
\label{e_bryan}
\end{equation}
where $x=\Omega_0(1+z)^3/\big(\Omega_0(1+z)^3+\Omega_{\Lambda_0}\big)-1$. Yet, it is much harder for a filament to virialise than for a halo, since it is a slender quasi-linear structure. Formed from the anisotropic collapse of matter along the axis of a cosmic web \citep{Peebles_1980}, filaments continue to be subject to tidal forces and ongoing accretion of matter even after their formation. Cosmic filaments will thus typically exist at most in a state of \textit{quasi-virialization} rather than full virialization. We also note that the virialization timescale of filaments is poorly understood.\par

For infinite self-gravitating filaments\footnote{Note that if the gas is immersed in the DM the gas filament is not self-gravitating.}, both isothermal hydrodynamic equilibrium (for gas) and the cylindrical steady-state solution of the collisionless Jeans equation (for DM) give a Plummer-like profile in the direction perpendicular to the filament axis \citep{Stodolkiewicz_1963},
\begin{equation}
\rho(r)=\frac{\rho_0}{\left(1+(r/r_0)^2\right)^2},
\label{Plummer}
\end{equation}
where the filament core radius reads $r_0 = \sqrt{2\mathcal{K}/(\pi G\rho_0)}$ and pressure $P=\mathcal{K}\rho$ is exerted by the transverse velocity dispersion $\mathcal{K}=\sigma_{\text{1D}}^2$ in the case of DM. In the limit where no external pressure is acting on the gas/DM filament, i.e. the gas/DM distribution in 3D is never cut off, the circumcylindrical integration boundary $R_b \rightarrow \infty$ and, upon integrating the circle surface out to $R_b$, the line mass per unit length $\zeta$  converges against the constant value \citep{Ostriker_1964}
\begin{equation}
\zeta = \frac{2\mathcal{K}}{G}.
\label{VirTheoremFils}
\end{equation}
Interestingly, one arrives at the same result \citep{Fiege_2000, Hennebelle_2013} by invoking the virial theorem $2K+W=0$ for a long DM filament of finite mass $M$ and length $L$. The gravitational potential energy $W$ and kinetic energy $K$ can be written as
\begin{equation}
W = -\frac{GM^2}{L}, \ K = \frac{1}{2}M(2\sigma_{\text{1D}}^2),
\end{equation}
and, upon setting the surface terms to zero, we again obtain Eq. \eqref{VirTheoremFils}. The reason why the two approaches yield the same result is that the virial theorem itself is based on the moment equations of the collisionless Boltzmann equation, i.e. Jeans equations, thereby not qualifying as an independent result. Since DM filaments form out of a fragmentation of their parent wall, the isolated filament profile in Eq. \eqref{Plummer} has to be corrected for this embedding. The starting point of this analysis is the density profile of a 2D sheet in equilibrium \citep{Miyama_1987}, either the hydrodynamic equilibrium for gas or the Jeans steady-state equilibrium for DM,
\begin{equation}
\rho(z) = \frac{\rho_{z_0}}{\cosh^2\left(z/z_0\right)},
\label{e_steady_state_sheet}
\end{equation}
where the scale height is given by $z_0 = \sqrt{\mathcal{K}/(2\pi G \rho_{z_0})}$.
The azimuthal average of this profile (i.e. a closed line integral in the plane of the sheet at height $z$) cannot be expressed in an analytically closed form, yet can be approximated by
\begin{equation}
\rho_{\text{corr}}(r) = \rho_{z_0}\frac{\tanh\left(\alpha r/z_0\right)}{\alpha r/z_0},
\label{e_steady_state_sheet_approx}
\end{equation}
with $\alpha \sim \pi/2$, see \cite{Ramsoy_2021}.\par 

Quasi-virialised cosmic filaments have two characteristic scales:
\begin{itemize}
\item As Eq. \eqref{VirTheoremFils} demonstrates, the gravitational energy of a DM filament will be compensated for by the kinetic energy of the DM particles, i.e. the transverse velocity dispersion, out to a certain distance from the filament spine. The distance where the velocity dispersion (or sound speed for gas) drops considerably is sometimes called the truncation radius $r_{\text{tr}}$ and encapsulates the breakdown of the hydrostatic model/steady-state Jeans approach. For DM, the truncation radius is somewhat analogous to the splashback radius for halos, as velocity dispersion can only be generated where shell crossing has occurred.
\item The second characteristic radius of cosmic filaments, the core radius $r_0$ from Eq. \eqref{Plummer}, specifies where the density drops off considerably. As a morphological approach, the NEXUS+ filament identification algorithm (cf. Sec. \ref{ss_nexus}) is sensitive to density variations and $r_0$, if resolved, is well recovered by NEXUS+ \citep{Cautun_2014}.
\end{itemize}

\section{Resolution Tests}
\label{s_res_tests}
\begin{figure*}
\begin{subfigure}{0.49\textwidth}
\includegraphics[scale=0.47]{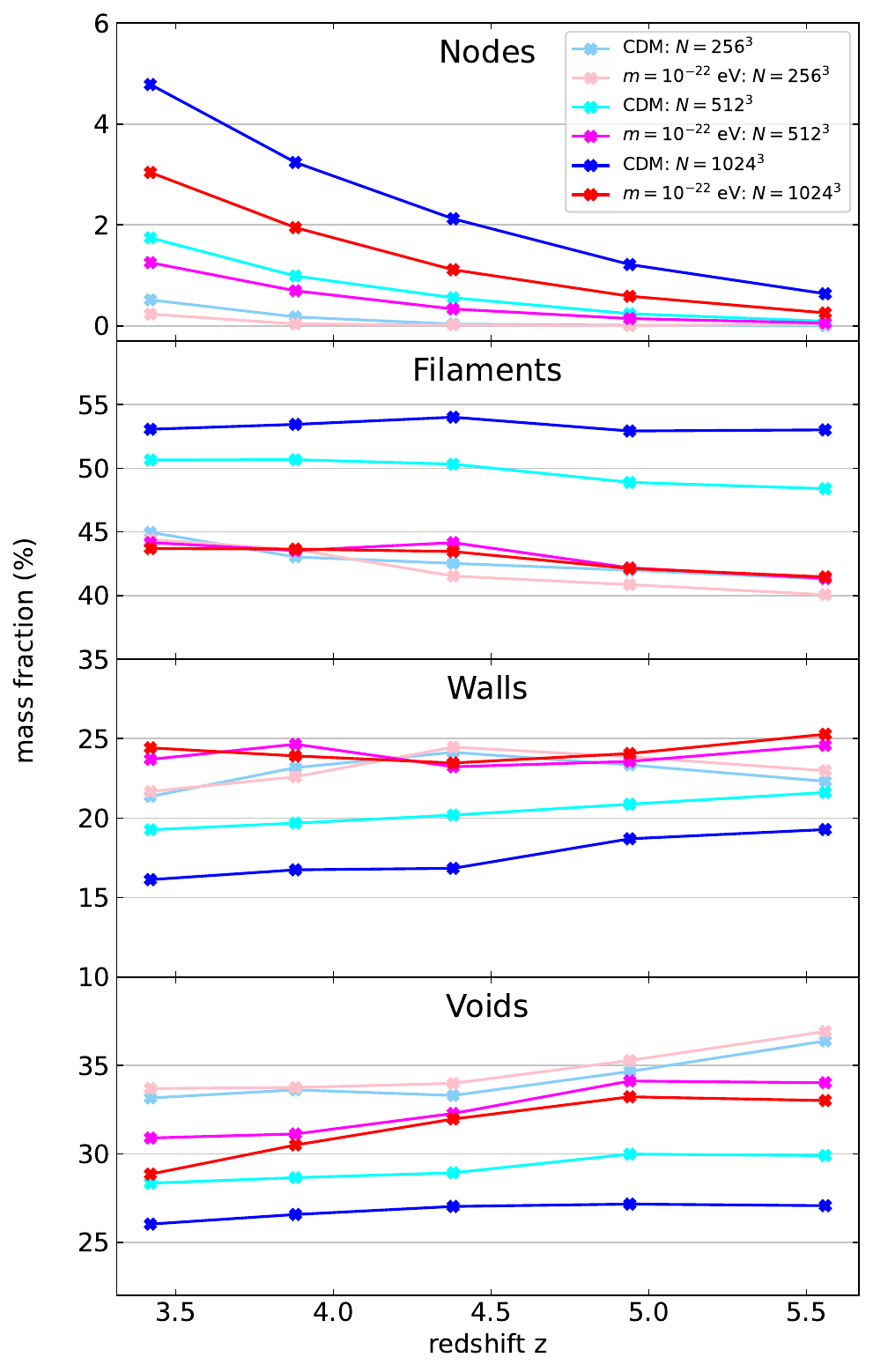}
\end{subfigure}
\begin{subfigure}{0.49\textwidth}
\includegraphics[scale=0.47]{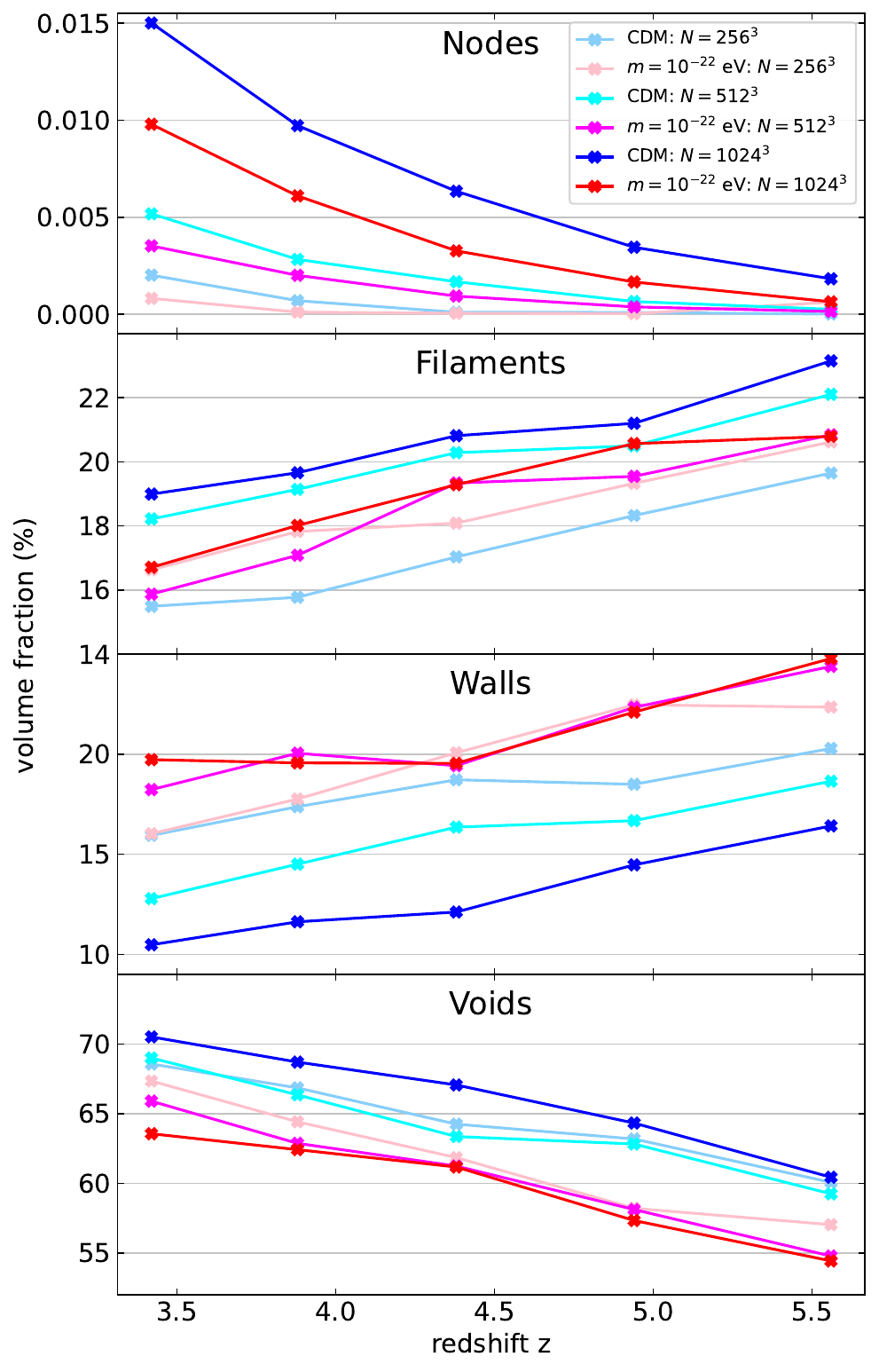}
\end{subfigure}
\caption{Evolution of the mass (left) and volume (right) filling fractions for the $N$-body CDM and cFDM (for $m=10^{-22}$ eV) runs across different resolutions $N=256^3, 512^3,1024^3$ and $L_{\text{box}} = 40\ h^{-1}$Mpc. Each row represents a different NEXUS+ cosmic web environment. Cosmologies and resolutions $N$ are differentiated by color as shown in the legend.}
\label{f_fracs_res}
\end{figure*}
\begin{figure}
\hspace{-0.3cm}
\includegraphics[scale=0.50]{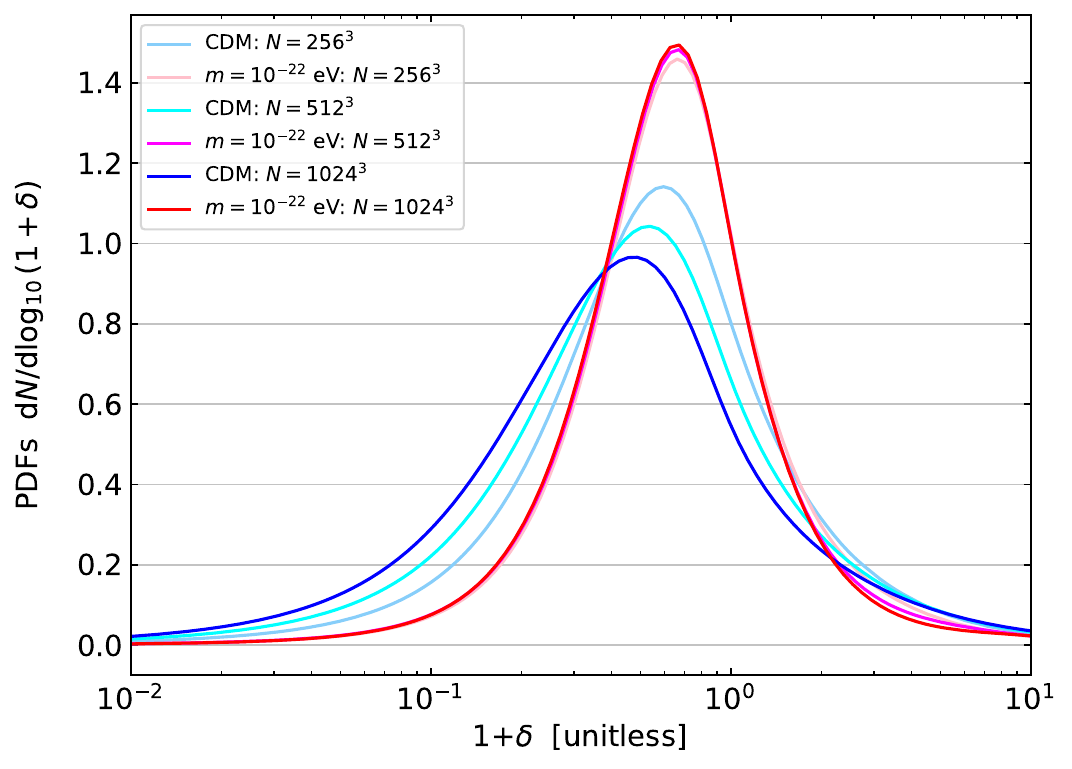}
\caption{Component-agnostic log overdensity PDFs for the $N$-body CDM and cFDM (for $m=10^{-22}$ eV) runs with varying resolutions $N=256^3, 512^3,1024^3$ and $L_{\text{box}} = 40\ h^{-1}$Mpc at redshift $z=3.9$. Cosmologies and resolutions $N$ are differentiated by color as shown in the legend.}
\label{f_dens_pdfs_tests}
\end{figure}
\begin{figure}
\hspace{-0.2cm}
\includegraphics[scale=0.49]{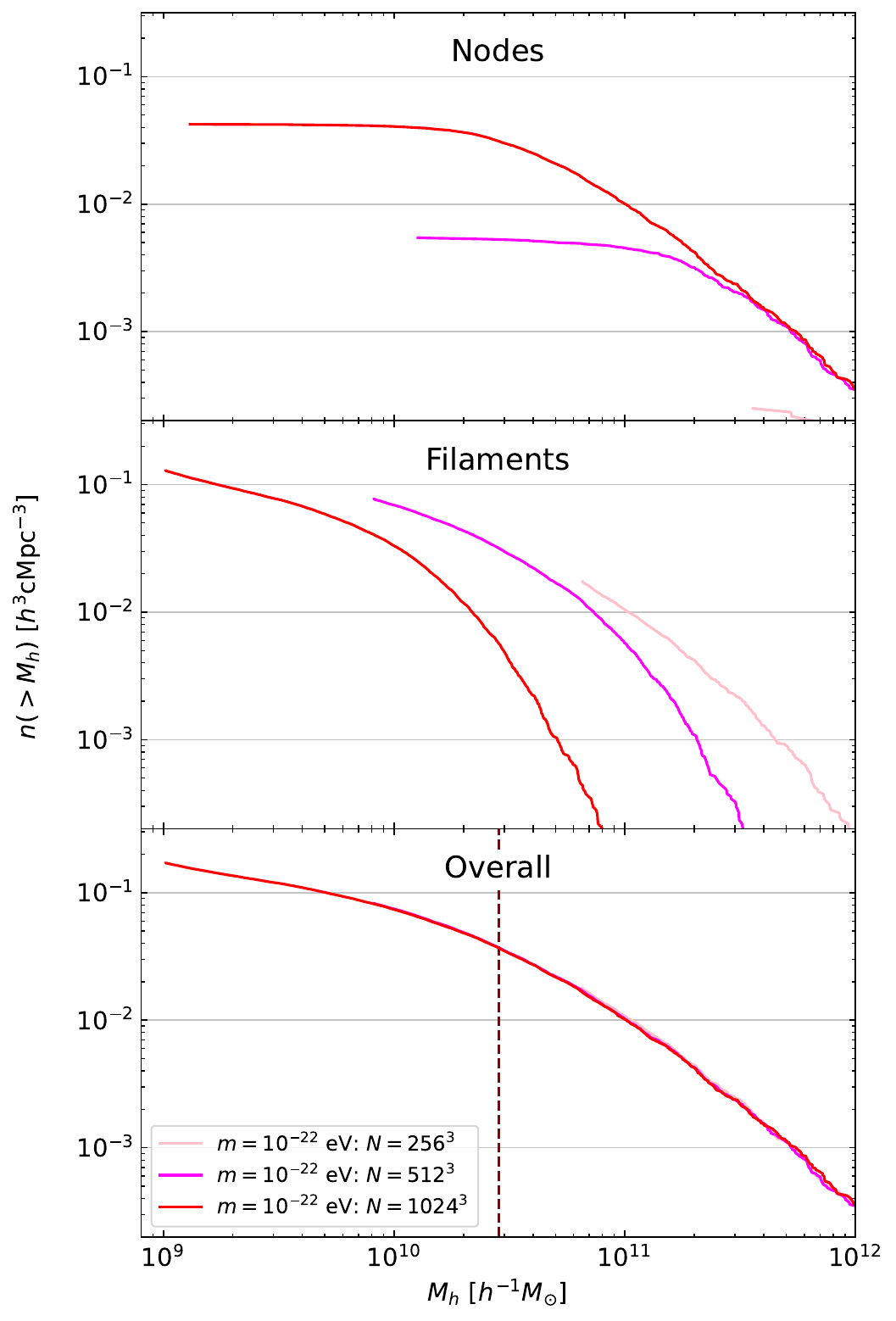}
\caption{Cumulative halo mass functions (cHMFs) for the $N$-body cFDM runs with $m=10^{-22}$ eV across three resolutions $N=256^3, 512^3,1024^3$ and $L_{\text{box}} = 40\ h^{-1}$Mpc at redshift $z=3.9$, split according to the two major NEXUS+ environments in which the halos reside, nodes (first row) and filaments (second row). The last row shows the overall cHMFs; the vertical dashed line denotes the half-mode mass $M_{1/2}$ of the $m=10^{-22}$ eV model.}
\label{f_hmfs_tests}
\end{figure}
The resolution dependence of the cosmic web statistics at high-$z$ is analysed in the following. The resolution or smoothing scale $\Delta x=L_{\text{box}}/N_{\text{lin}}$ can be thought of as a lens through which the cosmic web is observed, and a priori we do not expect its statistics to be invariant under changes of $\Delta x$, in the same way that the cosmic variance $\sigma^2$ and skewness $S_3$ depend on $\Delta x$. Since the cosmic web is close to being fractal on intermediate scales \citep{Gaite_2019}, there is some similarity to studying the resolution dependence of fractal characteristics for measured fractal-like morphologies such as contact interfaces or nanowires \citep{Swingler_2010, Liu_2018}.\par 

We run the NEXUS+ algorithm on the CDM and cFDM simulations of various DM resolutions $N=256^3$, $512^3$, and $1024^3$ (also used in the main text) and assess to what extent the morphological components depend (statistically averaged, not on a per-component basis) on the resolution scale $L_{\text{box}}/N_{\text{lin}}$. Note that the resolution studies of \cite{Cautun_2014} are slightly different in that they use the same simulation snapshots (from MS and MS-II), paint the matter density onto a regular grid of different resolutions and juxtapose the NEXUS+ statistics thereof. Especially for $N$-body simulations, the difference between the two approaches is small. However, since the resolution scale $\Delta x$ is a natural scale that follows directly from the simulation specifications, we feel that our resolution study approach is more ab initio.\\

\noindent\textbf{Mass and Volume Filling Fractions:}\newline
In Fig. \ref{f_fracs_res}, we show the resolution dependence of mass and volume filling fractions in CDM and the rather extreme cFDM model with $m=10^{-22}$ eV at redshifts\footnote{The $256^3$ and $512^3$ runs were only evolved down to $z=3.4$.} $z\sim 3.4-5.6$. We find \citep[as do][]{Cautun_2014} that volume filling fractions are less sensitive to the resolution scale than mass filling fractions, hence the following discussion focuses on the latter. The CDM trends identified in \cite{Cautun_2014} are recovered and extended to higher redshift. Node and filament mass fractions increase with increasing resolution $N$ while wall and void mass fractions decrease with $N$. Filament and void mass fractions exhibit the strongest dependence on the resolution scale with absolute differences between $N=256^3$ and $N=1024^3$ values partially exceeding $10$\%.\par

For all environments except for nodes, the relative difference between $N=512^3$ and $N=1024^3$ mass filling fractions ($\sim 3$\%) is about half of that between $N=256^3$ and $N=512^3$ ($\sim 6$\%). Disregarding cosmic variance stemming from a finite $L_{\text{box}} = 40\ h^{-1}$Mpc, a naive extrapolation would suggest that in order to obtain sub-$1$\% `convergence' in the extracted mass filling fractions at $z\sim 4$, a DM resolution of $N=4096^3$ would be necessary, corresponding to $\Delta x \sim 10 \ h^{-1}$kpc \citep[as opposed to $\Delta x \lesssim 0.4 \ h^{-1}$Mpc at $z\sim 0$,][]{Cautun_2014}. Since individual morphological components are highly anisotropic, the anisotropy-agnostic statement of CDM being scale-free on linear scales \citep[i.e. scales larger than $\Delta x_{\text{nl}} = 2\pi/0.14 \ (1+z)^{-2/(2+n_s)}\ \text{Mpc} \sim 15\ $Mpc at $z=3.9$,][]{Bull_2015} is not at odds with these conclusions. Mass and volume filling fractions thus tend to `converge' once $\Delta x$ becomes much smaller than the typical scale of cosmic environments.\par 
 
For cFDM, a reduced resolution dependence is observed for mass (and volume) filling fractions. In particular, in the cFDM model with $m=10^{-22}$ eV, the filament, wall and void mass as well as volume fractions vary only about $\sim 1-2$\% between $N=512^3$ and $N=1024^3$. This implies that for $m=10^{-22}$ eV cFDM at $z\sim 4$, a DM resolution of $N=2048^3$ would be sufficient to obtain sub-$1$\% `convergence' in the extracted mass and volume filling fractions.\par 

What causes the reduced resolution dependence of mass and volume filling fractions in cFDM? It is the lack of small-scale structure that renders the resolution dependence less significant, provided structure on scales larger than the cutoff scale $2\pi/k_{1/2}$ is captured by NEXUS+. Here, $k_{1/2}$ denotes the half-mode scale of the cFDM model \citep{Marsh_2016}. In other words, this behavior is expected as long as the resolution scale $\Delta x$ satisfies $2\pi/(\Delta x) > k_{1/2}$. This is naturally the case, however, since the small-scale cutoff in the primordial power spectrum is always resolved in the simulations \citep{Dome_2022}.\\

\noindent\textbf{Overdensity PDFs:}\newline
In Fig. \ref{f_dens_pdfs_tests}, we show log overdensity PDFs across various resolutions for CDM and $m=10^{-22}$ eV cFDM. The CDM log overdensity median migrates to smaller values as the resolution $N$ is increased, a trend predicted within the aforementioned framework of extreme value statistics \citep{Repp_2018} even if redshifts $z\sim 4$ and resolution scales $\Delta x \lesssim 0.15\ h^{-1}$Mpc are beyond its range of validity. Lower overdensity medians at higher resolution are also observed in other simulations \citep[e.g.,][]{Stucker_2018}. The PDF for $m=10^{-22}$ eV cFDM is more sharply peaked as we have seen in Sec. \ref{ss_overdens_pdfs} but, most importantly, shows hardly any resolution scale dependence. This is analogous to the reduced resolution scale dependence of mass and volume filling fractions in cFDM, cf. \mbox{Fig. \ref{f_fracs_res}}.\par 

The fast `convergence' of overdensity PDFs with $\Delta x$ in cFDM is very much at odds with results from analytical approaches devised to predict the matter distribution in scale-free cosmologies \citep{Klypin_2018, Repp_2018, Uhlemann_2020} and thus necessitates improved theories. Again, such weak resolution dependence is limited to $\Delta x$ values with $2\pi/(\Delta x) > k_{1/2}$.\\

\noindent\textbf{Halo Mass Distributions:}\newline
Globally averaged filament mass filling fractions in cFDM cosmologies have little dependence on the resolution scale $\Delta x$, especially for $m=10^{-22}$ eV (cf. Fig. \ref{f_fracs_res}, left panel). However, the filament-conditioned HMF does exhibit a strong dependence on $\Delta x$, as shown in Fig. \ref{f_hmfs_tests}. The smaller the resolution scale $\Delta x$, the fewer halos get identified as filament halos and the more as node halos. This is in agreement with the strong resolution dependence of node mass fractions in Fig. \ref{f_fracs_res}, which persists even for cFDM cosmologies. Node- and filament-conditioned HMFs depend on $\Delta x$ since especially the identification of nodes has a strong dependence on $\Delta x$. As more voxels get identified as nodes with the gradual decrease of $\Delta x$, node mass fractions increase (cf. Fig. \ref{f_fracs_res}, left panel) and node-conditioned HMFs increase at the expense of filament-conditioned HMFs (cf. Fig. \ref{f_hmfs_tests}).\par 

On the other hand, the overall environment-agnostic cHMF converges quickly with $\Delta x$. As $\Delta x$ decreases, smaller halo mass ranges of the cHMF get carved out while we find convergence for larger halo masses that are also captured at higher values of $\Delta x$. This behavior is expected since regardless of DM resolution the standard linking length of $b = 0.2 \times \text{(mean inter-particle separation)}$ adopted for the {\fontfamily{cmtt}\selectfont FoF} algorithm results in $\Delta \sim 180$ for an isothermal density profile \citep{More_2011}, which is close to the analytical spherical collapse-based result, Eq. \eqref{e_bryan}. Since virialisation of a halo is a physical property that is independent of $\Delta x$, we expect overall HMFs to converge with $\Delta x$.\par

\label{lastpage}
\end{document}